\documentclass{article}
\usepackage{nips,times}
\usepackage{amssymb}
\usepackage{url}
\usepackage{graphicx}
\usepackage{subfigure}
\usepackage{algorithmic}
\usepackage{common}
\title{TwitterCrowds:  Techniques for Exploring Topic and Sentiment in Microblogging Data}

\author{
Daniel Archambault \\
Department of Computer Science\\
Swansea University\\
\texttt{d.w.archambault@swansea.ac.uk} \\
\And
Derek Greene \\
School of Computer Science \& Informatics\\
University College Dublin\\
\texttt{derek.greene@ucd.ie} \\
\And
P\'{a}draig Cunningham \\
School of Computer Science \& Informatics\\
University College Dublin\\
\texttt{padraig.cunningham@ucd.ie} \\
}

\begin{document}

\maketitle
\begin{abstract}
Analysts and social scientists in the humanities and industry require techniques to help visualize  large quantities of microblogging data.  Methods for the automated analysis of large scale social media data (on the order of tens of millions of tweets) are widely available, but few visualization techniques exist to support interactive exploration of the results.  In this paper, we present extended descriptions of ThemeCrowds and SentireCrowds, two tag-based visualization techniques for this data.  We subsequently introduce a new list equivalent for both of these techniques and present a number of case studies showing them in operation.  Finally, we present a formal user study to evaluate the effectiveness of these list interface equivalents when comparing them to ThemeCrowds and SentireCrowds.  We find that discovering topics associated with areas of strong positive or negative sentiment is faster when using a list interface.  In terms of user preference, multilevel tag clouds were found to be more enjoyable to use.  Despite both interfaces being usable for all tested tasks, we have evidence to support that list interfaces can be more efficient for tasks when an appropriate ordering is known beforehand.
\end{abstract}

%-----------------------------------------------------
%!TEX root = twittercrowds-tr.tex
\section{Introduction}

Researchers in the humanities and social sciences, as well as many areas of private industry, require techniques to support the visualization of topics and associated sentiment in large scale microblogging data evolves over time. When the scale of the microblogging data becomes large, direct visualization or summaries of the data will not suffice, and it becomes necessary to employ machine learning techniques to extract meaningful signal from the data.  Particularly relevant is the area of unsupervised learning, where scalable clustering techniques can be used to summarize this data.  In the work presented here, we consider data sets in the range of tens of millions of tweets; these data sets are quite large when compared to those typically used in information visualization but relatively normal in the area of social media analysis.  

On the analysis side, however, there are still relatively few information visualization techniques available that are able to scale to data sets of this size.  Throughout the social media analysis literature, many algorithms have been proposed for uncovering trending topics \cite{becker11beyond,11wade}, and areas of strong sentiment \cite{Brew2011,tumasjan10predicting}.  However, visualization of this data is essential for hypothesis formation and explanation of this analysis.  Visualization can also assist in clarifying the reasons behind a trending topic or a particular area of strong sentiment.  For example, analysis of social media may reveal that a topic associated with {\it Japan} was trending in March 2011, but it would be difficult to automatically explain why it trended without involving visualization.  Also, it is unclear when and how the language around a topic evolves over time.  Automation can reveal time periods and topics of strong sentiment but cannot relate it to the type of language used around this sentiment all that easily.  As these types of questions are inherently imprecise, involving a human through visualization can help support answers to these questions.

When applying classification or cluster analysis techniques to microblogging data, it is common to take a user-centric approach~\cite{hannon10twitter}.  In this approach, posts generated by a user during a given time period (typically one day) are concatenated into a single ``profile document''. The data set for each day is then hierarchically clustered using one of a number of methods existing in the literature.  In information visualization, many techniques already exist to visualize hierarchically clustered data~\cite {09Elmqvist,10STAR}.  In order to allow for our visualizations to scale to the data set sizes required for this problem, we modify these information visualization techniques to explore the output of user-centric clustering approaches.

By developing these techniques, we hope to support a number of user tasks for large scale microblogging data.  Specifically, we would like to assist in helping users make sense of the language used around a given topic and how it evolves over time.  Second, given a topic cluster of particularly strong sentiment, we would like to create visualization interfaces to support understanding of the topics and language used in this area of strong sentiment.

In order to create effective visualization techniques for this type of data, we must consider a specific problem in the area of text visualization.  Currently, a standard visualization technique for understanding topics in documents is the tag cloud~\cite {viegas08tags}.  However, formal user experimentation has cast some doubt on their effectiveness for visualizing the theme of a document collection~\cite {07Rivadeneira} and for locating specific tags~\cite {07Halvey}.  In this work, we incorporate the findings of these experiments and extend this literature with a new experiment comparing the utility of tag clouds and lists in our visualization context.

%The primary contribution of this paper is a technique for visualizing large scale microblogging data that supports tasks involving understanding language usage around areas of strong sentiment and given topics.  
The primary contributions of this paper are list interface equivalents for ThemeCrowds and SentireCrowds and a formal evaluation of these four interfaces.  We present extended description and further case studies of the ThemeCrowds~\cite {11ThemeCrowds} and SentireCrowds~\cite {Brew2011} interfaces.  The paper provides examples of how all interfaces visualize the language used around topics and areas of strong sentiment in microblogging data through a number of case studies.  We subsequently evaluate, through a formal user study, the presented list interface, comparing it to the interfaces described in ThemeCrowds and SentireCrowds on the types of tasks we expect the interfaces to support.  In our experiment, we find that the discovery of topics associated with areas of strong positive or negative sentiment is faster when using a list interface.  In terms of user preference, multilevel tag clouds were found to be more enjoyable to use.  Despite both interfaces being usable for all tested tasks, our experiment provides support for the conclusion that list interfaces can be more efficient for tasks when an appropriate ordering is known beforehand.

\section{Related Work}
\label{sec:related}
Related work is divided into methods for microblogging data analysis and visualization.  We present both areas in this section.
%!TEX root = tvcg2013.tex
\subsection{Microblogging Data Analysis}

Microblogging services, such as Twitter, allow users to share content by posting frequent, short text updates. Many researchers have become interested in identifying and characterizing communities of users on Twitter,  sharing common interests and opinions. Java~\etal \cite{java07twitter}  initially demonstrated the presence of distinct Twitter user communities, where members share common interests as reflected by the terms appearing in their tweets. Kwak~\etal \cite{kwak10twitter} performed an evaluation based on a sample of 41.7 million users. The authors studied aspects such as: identifying influential users, information diffusion, and trending topics. Shamma~\etal \cite{09Shamma} performed an analysis of microblogging activity during the 2008 US Presidential debates, in terms of tweet content and user interactions. Unlike other text mining tasks, the authors noted that the informal and inconsistent use of vocabulary on Twitter made topic identification difficult. Becker~\etal \cite{tumasjan10predicting} performed content-based analysis of 100k tweets posted by users discussing politics in the lead up to the 2009 parliamentary elections in Germany. The authors examined both the degree of participation of individual political parties and the sentiment expressed by users towards the leaders of those parties. The potential for sentiment on Twitter to predict the outcome of an election was also discussed. Herda{\u{g}}delen~\etal \cite{herdag12news} discussed the formation of spontaneous topical groups on Twitter around news stories. These groups were formed by identifying sets of users sharing related news articles from The New York Times. 

Recently, a variety of researchers have focused on Twitter as a target for benchmarking sentiment analysis and opinion mining techniques. Pak \& Paroubek \cite{pak10twitter} collected Twitter data for this purpose and trained a \Naive Bayes classifier on both $n$-grams and part-of-speech tags to identify positive and negative tweets. Davidov \etal \cite{davidov10smileys} performed sentiment classification using different types of features, including punctuation, words, and $n$-grams.

%!TEX root = tvcg2013.tex
\subsection{Visualization}

A number of  systems have looked at ways of visualizing temporally-evolving textual data, some of which have been adapted to visualizing microblogging data.  ThemeRiver~\cite {themeRiver} encodes the frequency of terms as horizontal streams that grow and shrink over time.  D\"ork \etal \cite{10Dork} visualizes conversations in Twitter using a ThemeRiver-like approach.  Their system scales to data sets of over a million tweets and successfully identified conversations in the data.  Lee {\it et al.}~\cite {10Lee} presented a method that characterizes tags and their evolution in terms of frequency, by overlaying spark lines on each tag.  Rios and Lin~\cite {12Rios} analyze the evolution of events on Twitter using stream graphs, networks, and pixel oriented techniques.  In a similar way Wanner {\it et al}~\cite {12Wanner} present a shape-based representation for both topics and sentiment.  Best {\it et al.}~\cite {12Best} present a visual analytics system for themes and their prominence in Twitter data.  Huron and Fekete~\cite {12Huron} present a method for visualizing sentiment evolution over time and support a compromise between automated and manual sentiment analysis of Twitter data.  TextWheel~\cite {12Cui} provides a system for visualizing key terms and their emergence in large scale news streams.  The system also provides functionality for visualizing sentiment in these terms as well.

In a similar way, a number of  systems have looked at visualizing document clusters and how they evolve over time.  IN-SPIRE~\cite {95Wise} creates landscapes of documents using dimensionality reduction based on document statistics.  Hetzler {\it et al.}~\cite {05Hetzler} use animation to depict dynamically evolving  clusters and their system has facilities to take snapshots of the data over time.  Rose {\it et al.}~\cite {09Rose} summarize the evolution of collections of news stories and the topics they discuss. In their system, stories are clustered into their most highly associated theme and the system shows how the stories and keywords evolve over time.  Shi {\it et al.}~\cite {10Shi} combine trend graphs with tag clouds to  visualize cluster content and size as it evolves over time. 
Cui {\it et al.}~\cite {11Cui} present a system for visualizing the emergence, split, and merge of themes in time-stamped documents.  The technique is primarily built around stream graphs and uses critical events as a basis for visualization.  Whisper~\cite {12Cao} visualizes the diffusion of information over time in microblogging data from a geospacial perspective.

A number of techniques support the investigation of documents clustered at multiple levels of resolution.  InfoSky~\cite {04Granitzer} provides a way to visualize hierarchically clustered documents using a telescope metaphor.  FacetAtlas~\cite {10Cao} supports the visualization of unstructured text data at multiple levels of resolution, but focuses on depicting facets and the relationships between entities in document collection.

The above systems visualize temporal and/or topic in document collections.  However, these techniques do not support the visualization of dynamic, hierarchically clustered data.  The ThemeCrowds~\cite{11ThemeCrowds} and SentireCrowds~\cite {Brew2011} techniques, discussed in section~\ref {sec:methods-viz}, support the visualization of hierarchically clustered, dynamic twitter data.  In this paper, we introduce a list version of both techniques (section~\ref {secListInterface}) and evaluate its effectiveness through a formal user study (section~\ref {sec:eval2}).

A number of user studies have been run to test the effectiveness of tag clouds when compared to ordered lists of words.  Rivadeneira {\it et al.}~\cite {07Rivadeneira} tested tag clouds and lists where the words were ordered by frequency and alphabetically.  The study found that lists ordered by frequency provided the best performance in terms of understanding the theme of a single document. Halvey and Keane~\cite {07Halvey} performed an experiment comparing lists and tag clouds to find specific key terms.  In their experiment, the authors found that lists ordered alphabetically performed the best.  Our system and experiment are informed by these studies.  As we are primarily interested in tasks that involve theme, we order the terms in our lists and tag clouds according to frequency.  Our experiment differs from these studies in that it tests the visualization of clusters of user, microblog profiles centred around specific topics and areas of strong sentiment.

\section{Methods}
\label{sec:methods}
In this section, we describe our data analytics pipeline and visualization techniques for topics and sentiment in microblogging data.
%!TEX root = twittercrowds-tr.tex

\subsection{User Clustering}
\label{sec:methods-cluster}

The visualization systems proposed in this paper take a time series of multilevel clusterings of Twitter users as their input, where a clustering represents a snapshot of discussions on Twitter for a fixed {\bf{time~step}} (\eg a 24 hour period). Due to the size of the data sets used in the evaluations presented in this paper, we employ a scalable version of min-max linkage agglomerative hierarchical clustering (AHC)\cite{ding02cluster}, which consists of three distinct phases.

\noindent \emph{Phase 1}: We employ a ``problem decomposition'' strategy similar to the fractionation approach proposed in \cite{cutting92scatter}. This involves dividing a full data set of $n$ documents into $p$ smaller fractions of size $\approx n/p$, based on the random assignment of documents to fractions. Each fraction is clustered using min-max AHC and cosine similarity,  until a set of $k_{low}$ low-level leaf clusters have been identified. For each leaf cluster, we calculate its centroid vector. After all fractions have been clustered, the data set is now represented by $n' = p \times k_{low}$ {\bf{compressed vectors}} denoted $\fullset{v}{n'}$, where $n' \ll n$. Each vector $v_{i}$ corresponds to the centroid of a cluster $C_i$ identified in a fraction of the data containing $n/p$ documents, and will have a weight proportional to the cluster size $\abs{C_i}$:
\[
w_{i} = \frac{\abs{C_i}}{ n/p }
\]
Weights take a value $\in [0,1]$, such that compressed vectors derived from larger clusters will receive larger weights.

\vskip 0.8em
\noindent \emph{Phase 2}: We apply min-max agglomerative clustering to the set of $n'$ compressed vectors from Phase 1. In this case we use a modification of the linkage function that takes into account the weights assigned to the vectors: we replace cosine similarity in AHC with weighted cosine similarity, which is defined for a pair of compressed vectors $(v_{i},v_{j})$, with weights $w_{i}$ and $w_{j}$ respectively, as:
\begin{equation}
\label{eqn:wcos}
wcos(v_{i},v_{j}) = w_{i} \; w_{i} \; cos(v_{i},v_{j})
\end{equation}
At the start of the agglomeration process, each compressed vector $v_{i}$ is assigned to a singleton cluster. Pairs of clusters are repeatedly merged until a single node remains in the hierarchy. 

\vskip 0.8em
\noindent \emph{Phase 3}: Based on the clustering of compressed vectors from Phase 2, we build a hierarchical clustering for the $n$ original documents as follows. Firstly, the tree produced from Phase 2 is cut off at a point which yields $k \leq n'$ leaf clusters, and the centroid vectors for these clusters are calculated. A higher value for $k$ will yield a deeper hierarchy. Then, for each of the $n$ documents, we apply a nearest centroid classification procedure which assigns the document to the cluster with the most similar centroid. The resulting $k$ clusters form the leaf nodes of the complete tree. Using these leaf node assignments and the corresponding parent-child relations from the hierarchy generated in Phase 2, we subsequently reconstruct a complete tree for the original data set. 

\vskip 0.8em
To cluster users based on the content of their tweets, for every user we create {\bf{user profile}} documents ~\cite{hannon10twitter}, each containing the concatenation of all their tweets published during a given time step. The agglomerative clustering algorithm described above is applied to the user profiles to generate cluster hierarchies for each time step.  To identify the descriptive tags for each cluster in the hierarchy, we select the highest-weighted terms from the cluster's centroid vector.

\subsection{Sentiment Analysis}
\label{sec:methods-sent}

Once a hierarchical clustering of users has been generated, we calcculate sentiment scores for all nodes in the hierarchy using an approach analogous to the macro-level approach used previously in \cite{Brew2011}, which is based on the ``happiness index'' proposed by Kramer~\cite {kramer:happiness} for the analysis of Facebook data. As with clustering, sentiment scoring is performed on user profile documents in each time step. For each document assigned to a cluster, we count the frequency of positive and negative sentiment terms. These counts are normalized with respect to the mean and standard deviation of the positivity and negativity scores. The cluster sentiment score is calculated as the average score of all profiles assigned to that cluster:
\begin{equation}
H(C) = \frac{\mu_{pc} - \mu_{p}}{\sigma_{p}} - \frac{\mu_{nc} - \mu_{n}}{\sigma_{n}}
\label {happynessIndex}
\end{equation}
where $\mu_{ic}$ represents the fraction of terms that were positive $(i=p)$ or negative $(i=n)$ for cluster $C$, averaged across every document in the cluster. $\mu_p$ and $\mu_n$ are the average positive and negative word counts for all documents in the overall data set, and $\sigma_p$ and $\sigma_{n}$ are the corresponding standard deviations. 

The normalization in Equation~\eqref{happynessIndex} ensures that the positivity and negativity scores both contribute in a balanced way to the overall sentiment score.  Thus, a rise in sentiment may not only be due to increased positive term usage, but could also be the result of a drop in negative term usage.

%One subtle difference between our proposed micro- and macro-level scoring mechanisms is that, due to the use of an unweighted average, users who tweet often do not contribute more to the overall sentiment of the cluster.
%!TEX root = twittercrowds-tr.tex
\subsection{Visualization}
\label{sec:methods-viz}
In this section, we describe the ThemeCrowds~\cite {11ThemeCrowds} and SentireCrowds~\cite {Brew2011} visualization techniques.  Then, we describe the list interface equivalent to these interfaces.  

\subsubsection {ThemeCrowds}
\label {secThemeCrowds}

\begin {figure}[!t]
\centering
\subfigure [Topic Antichain] {\label {figAntichain}\includegraphics[width=0.45\linewidth] {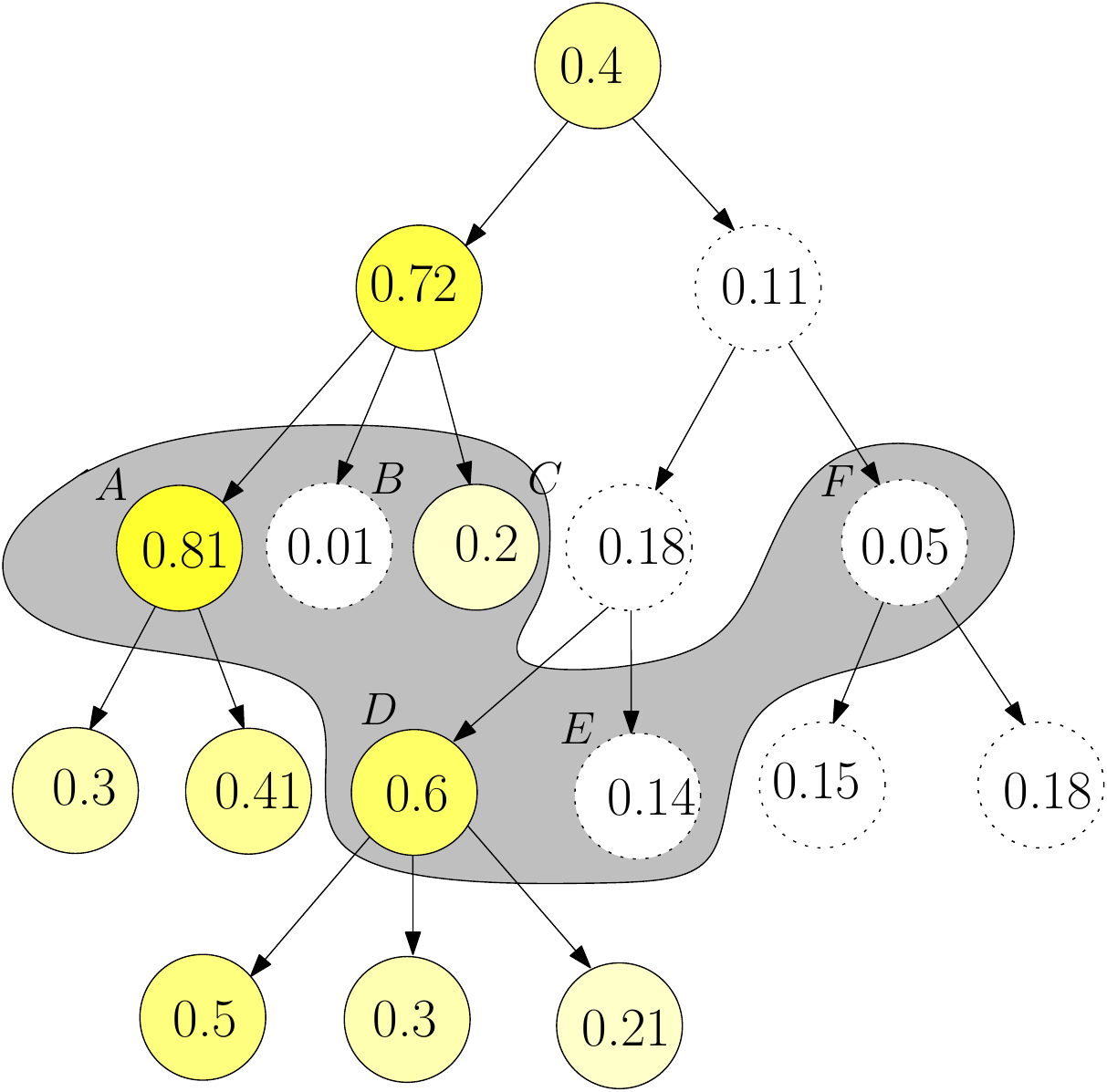}}
\hskip 4em
\subfigure [Multilevel] {\label {conversionToMulti}\includegraphics[width=0.38\linewidth] {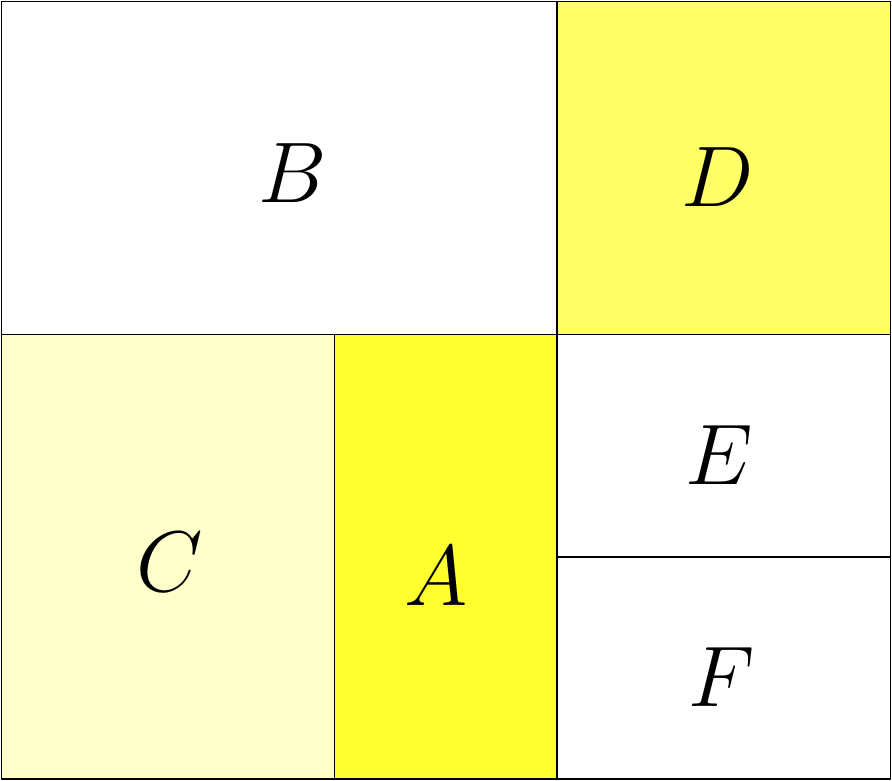}}
\caption {Selecting the antichain by topic automatically and converting to a multilevel tag cloud.  {\bf (a)} Automatic maximal antichain is selected such that each path in the hierarchy is cut exactly once and each node on the antichain subtends a subtree whereby all descendants have smaller match scores.  {\bf (b)}  Nodes on the antichain are converted to a multilevel tag cloud representation.  Each box represents a node in the hierarchy and the keywords of the discussion are presented as a tag cloud inside the treemap cell.  The order, from strongest to weakest match score is: $A, D, C, E, F, B$.}
\label {figThemeCrowdsMulti}
\end {figure}

\begin {figure}[!t]
double {\bf findMaxAntichain} ($r$)
\begin {algorithmic}
\STATE $m_v \leftarrow -1$
\FOR {$\forall c \in$ children of $r$}
\STATE $c_v \leftarrow$ {\bf findMaxAntichain} ($c$)
\IF {($c_v > m_v$)}
\STATE $m_v \leftarrow c_v$
\ENDIF 
\ENDFOR
\IF {($r_v > m_v$) or ($m_v < \theta$ and $r_v < \theta $)}
\STATE coarsen antichain to $r$
\STATE {\bf return} $r_v$
\ELSE
\STATE {\bf return} $m_v$
\ENDIF
\end {algorithmic}
\caption {Algorithm to find the best matching antichain.  The match score 
for each node in the tree is computed beforehand and is supplied as input.
The current root of the subtree is $r$ and its match score is $r_v$.  The maximum match score for any node in the subtree rooted at $r$ is $m_v$.  The value $\theta$ is the match threshold (everything below $\theta$ is considered as zero).  All nodes present on the antichain are the crowds of coarsest resolution that
have maximal match scores when compared to all nodes in the subtrees they subtend.}
\label {figBestMatchAlg}
\end {figure}

ThemeCrowds~\cite {11ThemeCrowds} is a technique for visualizing Twitter data that has been processed using the clustering method described in section~\ref {sec:methods-cluster}.  The visualization interface is designed to depict how the language around a given topic evolves over time.

Fig.~\ref {dncMulti} shows the ThemeCrowds interface.  The user can type a topic keyword or hashtag in the search box or select a cluster to find similar clusters.  After hitting return, an appropriate level of resolution is computed for the hierarchy.  A scented widget~\cite {07Willett} displays the magnitude of the discussion around the topic over time, with the height of the stream graph indicating the number of users involved in the discussion on a given day.  A black scroll bar indicates the time window visible in the multilevel tag cloud display showing evolution in language around the topic.  A small multiples~\cite {90Tufte} shows the six days of the visible time window indicated by the scented widget.  In the original version of ThemeCrowds, each day was encoded using a multilevel tag cloud.  The size of each box corresponded to the number of users discussing the topic and saturation indicating the relevance to the searched topic.

A {\bf maximal antichain} computation determines the appropriate level of resolution in the hierarchy.  Maximal antichains cut every path to the leaves of the hierarchy exactly once and have been used for graph visualization previously~\cite {09Elmqvist,10STAR}.  As we only deal with maximal antichains in this paper, we refer to them as {\bf antichains}.  Before the process of computing an antichain, every node in the hierarchy is assigned a {\bf match score}.  Let $r$ be a node in the hierarchy associated with a cluster of users.  If a search term is entered, this score is the ratio between the frequency of the search term in $r$ to the maximum frequency of the search term in any cluster of the data set.  If a cluster is selected, the score is the cosine similarity between $r$ and the selected cluster.  

Figs.~\ref{figThemeCrowdsMulti}~and~\ref {figBestMatchAlg} show our approach to computing this antichain given a hierarchy with match scores.  Our method assumes that the best matches should be displayed and given matches of equivalent relevance the coarsest level of the hierarchy should be displayed.  Let $\theta$ be a threshold match score whereby all values below this score are considered as $0$.  This score is used to decrease sensitivity to noise for our approach and is set to $0.2$ in the default implementation of ThemeCrowds.  The algorithm recursively traverses the hierarchy bottom-up in order to compute the antichain.  At a given node $r$ in the traversal with match score $r_v$, the algorithm places $r$ on the antichain if $r_v$ is greater than the match score of all of nodes in the subtree that it subtends (the first condition).  However, $r$ can also be placed on the antichain if $r$ and all of the nodes in the subtree it subtends have match scores less than $\theta$ (the second condition).  If either of these conditions is met, $r_v$ is returned and $r$ is placed on the antichain.  Otherwise, the current antichain remains unchanged and $m_v$, the current maximum match score of the subtree, is returned.

In the original version of ThemeCrowds, the nodes on the antichain are converted to a multilevel tag cloud as shown in Fig.~\ref {figThemeCrowdsMulti}.  Each box represents a node lying on the antichain of the hierarchy and the keywords of the discussion are presented as a tag cloud inside the treemap cell.  Words in the tag cloud are ordered by frequency~\cite {07Rivadeneira} as they are good for tasks involving uncovering the theme of a cloud.  The size of the box encodes the number of users in the cluster and saturation encodes the match score.  The layout of the tree map is computed using the Tulip~\cite {gdauber} implementation of the squarified treemap algorithm of Brules {\it et al.}~\cite {00Bruls}.

\subsubsection {SentireCrowds}
\label {secSentireCrowds}
\begin {figure}[!t]
\centering
\subfigure [Sentiment Antichain] {\label {figSentCut}\includegraphics[width=0.45\linewidth] {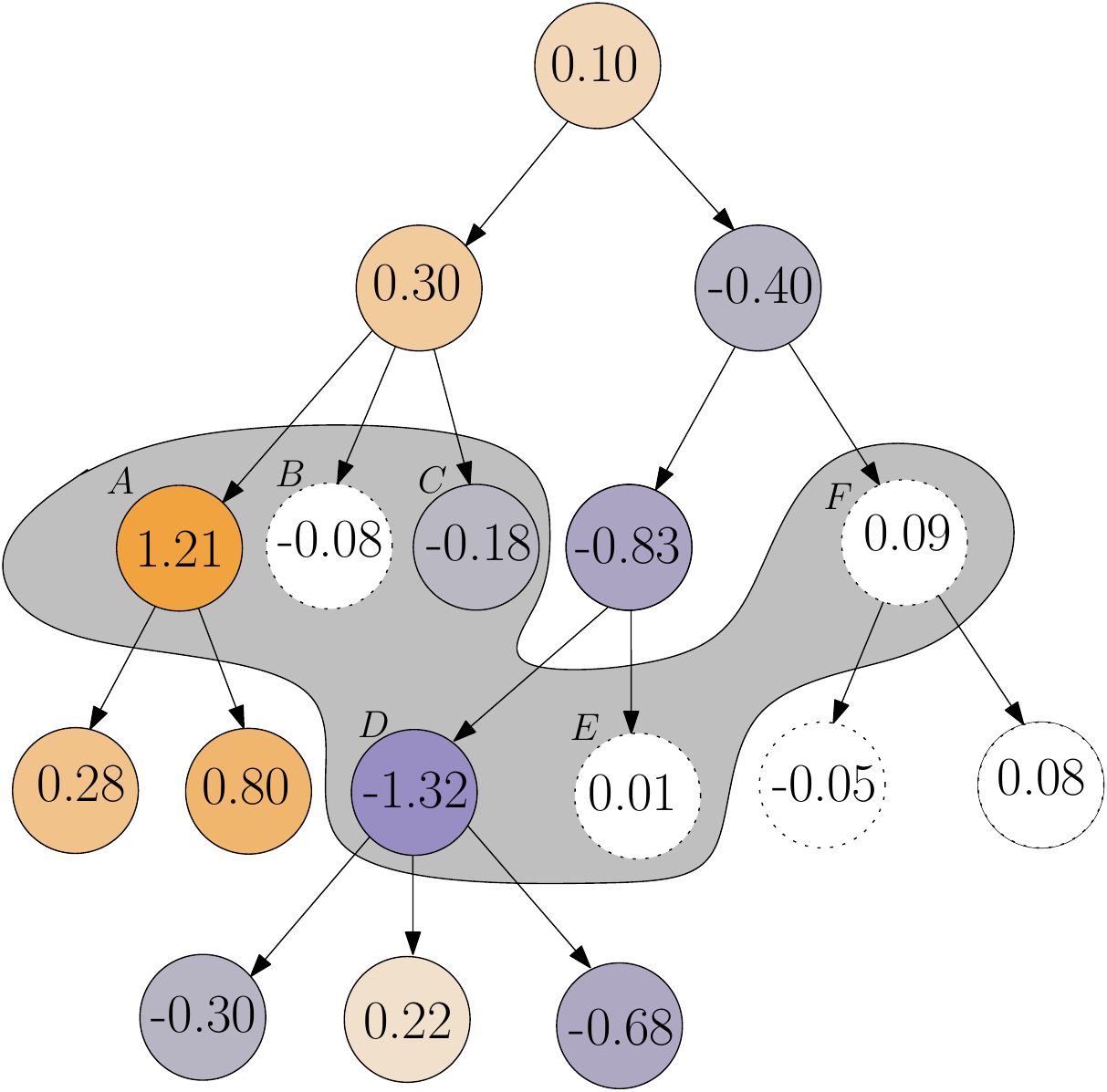}}
\hskip 4em
\subfigure [Multilevel] {\label {figSentAntichainMulti}\includegraphics[width=0.38\linewidth] {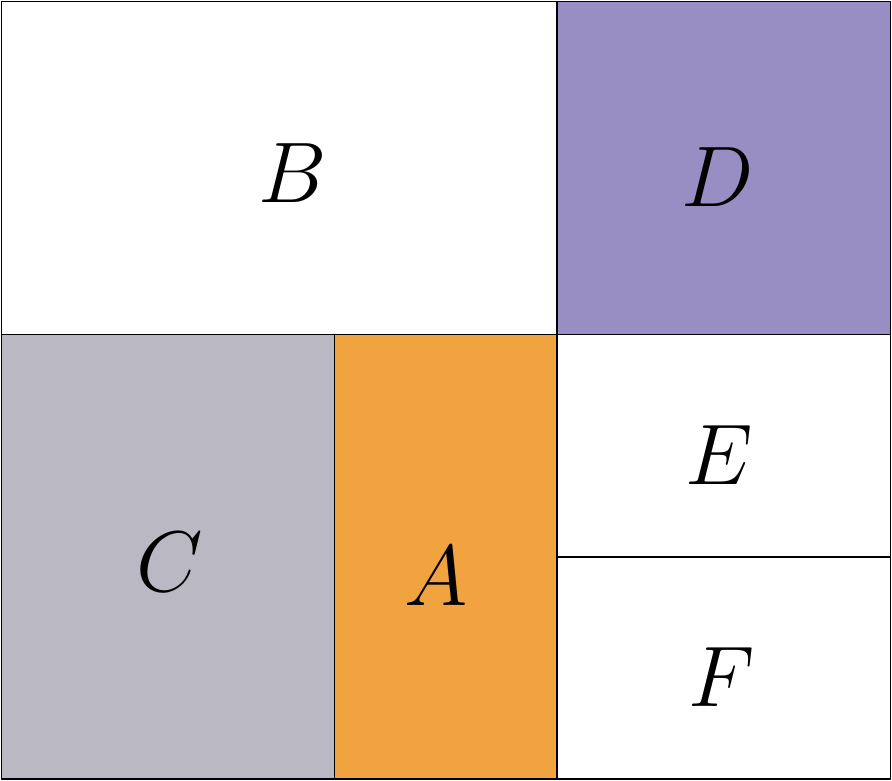}}
\caption {Selecting the antichain by sentiment automatically and converting it to a multilevel tag cloud.  {\bf (a)} Automatic maximal antichain is selected such that each path in the hierarchy is cut exactly once and each node on the antichain subtends a subtree whereby all descendants have more neutral sentiment scores (can be substituted less positive or less negative).  {\bf (b)}  Nodes on the antichain are converted to a multilevel tag cloud representation.  The hierarchy is computed exclusively based on topic and the antichain is computed based on sentiment.  Tag clouds and user clusters are determined in the same way as ThemeCrowds (section~\ref {secThemeCrowds}).  The order, from strongest to weakest, sentiment score is:  $D, A, C, F, B, E$.  Fig.~\ref {listRep} shows an equivalent list representation for this antichain and hierarchy.}
\label {figSentireCrowdsMulti}
\end {figure}

SentireCrowds~\cite {Brew2011} is a technique for visualizing topic and sentiment simultaneously in Twitter data that has been processed using the clustering method described in section~\ref {sec:methods-cluster}.  It is inspired by the ThemeCrowds approach presented in the previous section.

Fig.~\ref {easterRoyalMulti} shows the SentireCrowds interface.  The scented widget now encodes the maximum positive (tan stream graph) and negative (purple stream graph) sentiment score on a given day.  Antichains can be computed for positive sentiment only (``positive mode''), negative sentiment only  (``negative mode''), or both positive and negative sentiment  (``positive/negative mode'').  The remaining examples in this paper show SentireCrowds operating in positive/negative mode.

As with ThemeCrowds, SentireCrowds starts with a hierarchical clustering of users, grouped by text similarity, which encodes groups of users that are discussing similar topics on a given day.  In contrast to ThemeCrowds, a sentiment score is associated with each cluster of the hierarchy rather than a match score. Sentiment scores are calculated using Equation~\eqref{happynessIndex}.

%.  This sentiment score is based on the happiness index of Kramer~\cite {kramer:happiness}:
%%
%\begin{equation}
%H_c = \frac{\mu_{pc} - \mu_{p}}{\sigma_{p}} - \frac{\mu_{nc} - \mu_{n}}{\sigma_{n}}
%\label {happynessIndex}
%\end{equation}
%%
%In Equation~\eqref {happynessIndex}, $\mu_{ic}$ represents the percent of terms that were positive $(i=p)$ or negative $(i=n)$ for cluster $c$, averaged across every tweet collected in the cluster. $\mu_p$ and $\mu_n$ are the average positive and negative word counts for all tweets in the data and $\sigma_p$ and $\sigma_{n}$ are the standard deviations. This approach allows positivity and negativity scores to be normalized so that each score contributes in a balanced way to the overall happiness score.  Thus, a rise in the happiness score may not only be due to increased positive term usage, but could also be due to a drop in negative term usage.

In SentireCrowds, antichains are computed using the same algorithm described in Fig.~\ref {figBestMatchAlg} but using the absolute value of the sentiment scores from Equation~\eqref {happynessIndex}, when operating in positive/negative mode.  For positive mode, the sentiment score is used directly while for negative mode the negative of the sentiment score is used.  Fig.~\ref {figSentireCrowdsMulti} shows how these antichains are converted to multilevel tag clouds.  Positive sentiment is indicated in tan while negative sentiment is indicated in purple.  Neutral sentiment is indicated with white and saturation is used to encode how positive or negative a cluster is relative to neutral sentiment.  The layout of the tree map is computed using the Tulip~\cite {gdauber} implementation of the squarified treemap algorithm of Brules {\it et al.}~\cite {00Bruls}.

\subsubsection {Equivalent List Interfaces}
\label {secListInterface}

\begin {figure}[!t]
\centering
\subfigure [Sentiment Antichain] {\label {figSentAntichain}\includegraphics[width=0.45\linewidth] {figures/autoAntichainSent}}
\hskip 5em
\subfigure [List] {\label {figConversionToList}\includegraphics[width=0.17\linewidth] {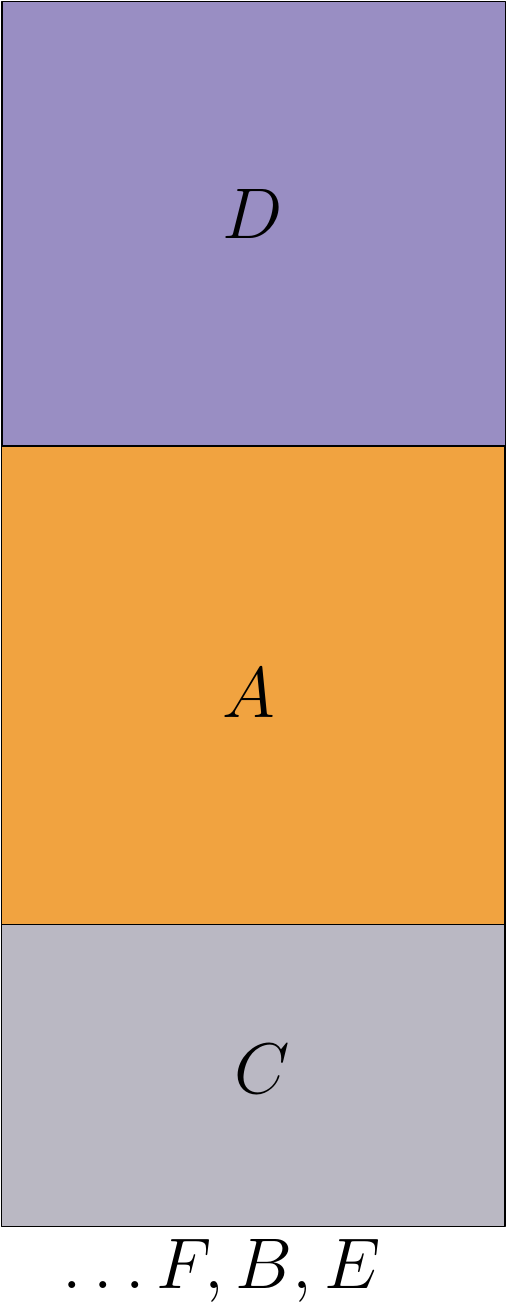}}
\caption {Selecting the automatic antichain and converting it to a list.  {\bf (a)}  Automatic maximal antichain is selected as in Fig.~\ref {figSentCut}. {\bf (b)}  Nodes on the antichain are converted to a list representation.  Each box in this representation corresponds to a node in the hierarchy and the most frequent keywords are presented in frequency order.  The nodes are sorted via sentiment deviation from neutral sentiment score: $D, A, C, F, B, E$.  The antichain and hierarchy are exactly the same as the multilevel tag cloud representation in Fig.~\ref {figSentireCrowdsMulti}.}
\label {listRep}
\end {figure}

The list interface, introduced in this paper, is based on the the ThemeCrowds and SentireCrowds techniques, but instead of presenting the entire hierarchy as a treemap, the groups of users are presented as items of a list.  Recent human computer interaction experiments have indicated that lists ordered by frequency outperform tag cloud presentation methods for understanding topics~\cite {07Rivadeneira} and locating specific tags~\cite {07Halvey}.  A description of how this interface represents maximal antichains is presented in Fig.~\ref {listRep}.  This interface is shown in action as ThemeCrowds in Fig.~\ref {osamaList} and as SentireCrowds in Fig.~\ref {consulateList}.

Fig.~\ref {listRep} is the list equivalent of Fig.~\ref {figSentireCrowdsMulti}.  In the list representation, each cluster of the multilevel hierarchy is first associated with a match or sentiment score.  Both colour encoding and antichain selection are done in the exact same way as previously described in ThemeCrowds and SentireCrowds.  The list of clusters is ordered according to match or sentiment (absolute value) score with more relevant or stronger sentiment conversations appearing at the top of the list.  Within each cluster, a list of keywords about the conversation is presented, ordered by frequency to facilitate comprehension of topic~\cite {07Rivadeneira}.  Words do not have a size proportional to frequency as they would in tag cloud representations.  Size no longer encodes the number of users in the cluster.  Instead, a label at the top of the cluster, {\it U: $\mathit{<}$number$\mathit{>}$}, reports how many users are present in the cluster.

%!TEX root = twittercrowds-tr.tex
\section {Case Studies}
\label{sec:eval1}

We tested our visualization techniques and above-described clustering method on two Twitter data sets.  Figs.~\ref {figUSCities}~and~\ref {figElection} show example results.  In this section, we describe the data sets and their clustering in section~\ref {sec:data} and our qualitative visual results in section~\ref {sec:discussion}.

\subsection{Data sets}
\label{sec:data}

Two Twitter data sets were used for our experiment and the qualitative results presented in this paper.  These data sets were both of realistic size and complexity for tasks in microblogging content analysis.

The {\tt US~Cities} data set, originally described by Brew {\it et al.}~\cite{Brew2011}, consists of 12,781,243 tweets from 336,802 unique users collected in 2011. We divided this corpus into 82 non-overlapping, 24-hour time steps. A multi-lingual, stop-list filter was applied to remove non-content-bearing terms. We also removed Twitter username mentions and URLs.  For each time step, we constructed profiles for all users active during that time period -- on average each time step contained $\approx 24k$ unique users. We applied the   clustering algorithm described in \refsec{sec:methods-cluster} to the resulting user profiles for each time step, where the time step data was divided into $p=5$ fractions and $k=50$ leaf nodes were generated to provide deep hierarchies.

Following previous interest in analyzing political activity on Twitter \cite{tumasjan10predicting}, we collected the {\tt Election~2012} data set, which consists of 1,253,055 tweets collected during the 2012 US Presidential Election, covering the period from 1st August to 31st October. The tweets were posted by a curated list of 2,404 users, which includes US politicians, political organizations, media outlets, and journalists. We divided the data set into 90 non-overlapping daily time steps, and pre-processed the time step data as above. This resulted in each time step having an average of 1,315 unique users. For this smaller data set, during the clustering process, the data was divided into $p=2$ fractions, and hierarchies with $k=30$ leaf nodes were generated for each step.

\subsection{Discussion}
\label{sec:discussion}
\begin {figure*}
\centering
\subfigure [Osama bin Laden (ThemeCrowds)] {\label {osamaList}\includegraphics[width=\linewidth] {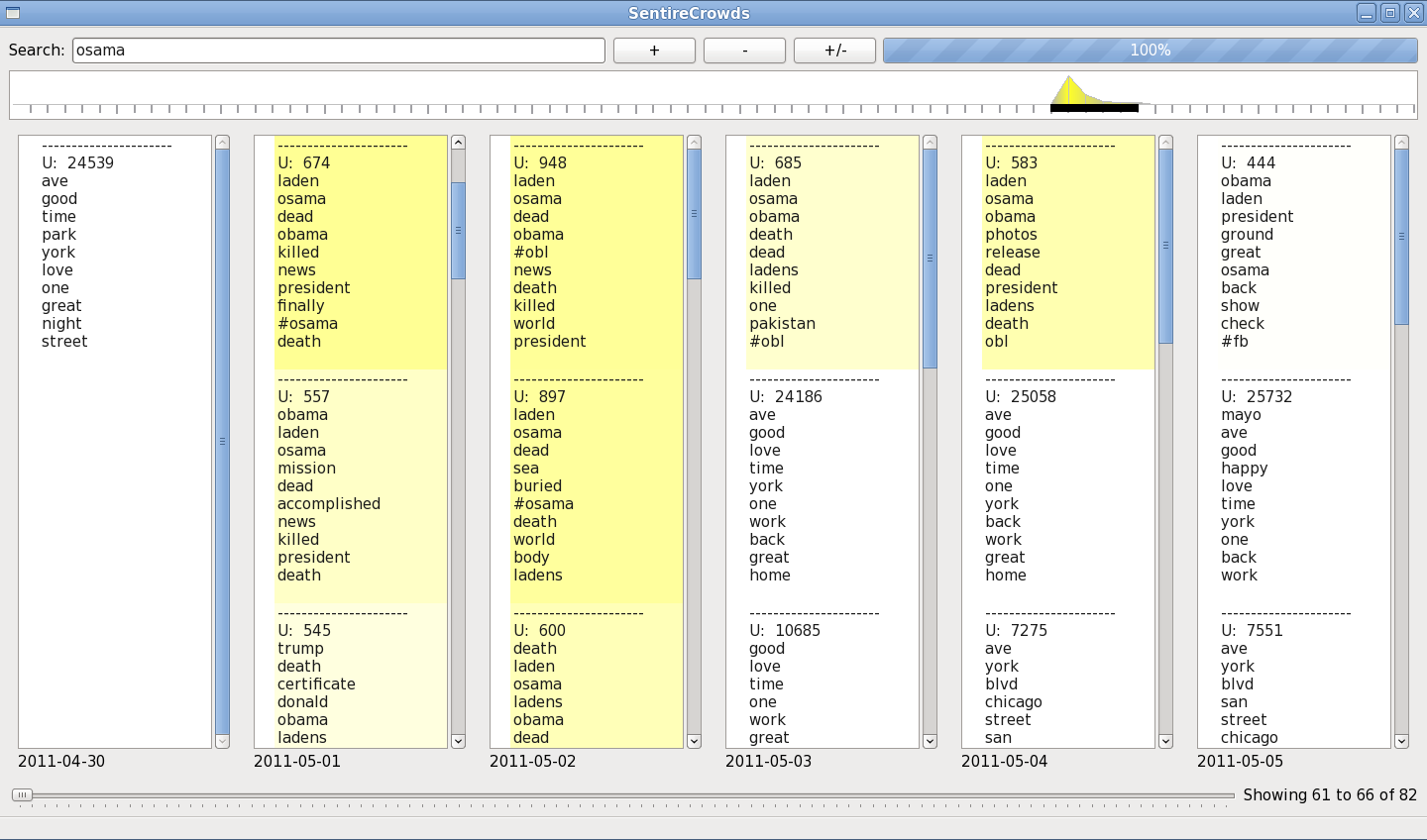}}
\vskip 1em
\subfigure [Easter and Royal Wedding (SentireCrowds)] {\label {easterRoyalMulti}\includegraphics[width=\linewidth] {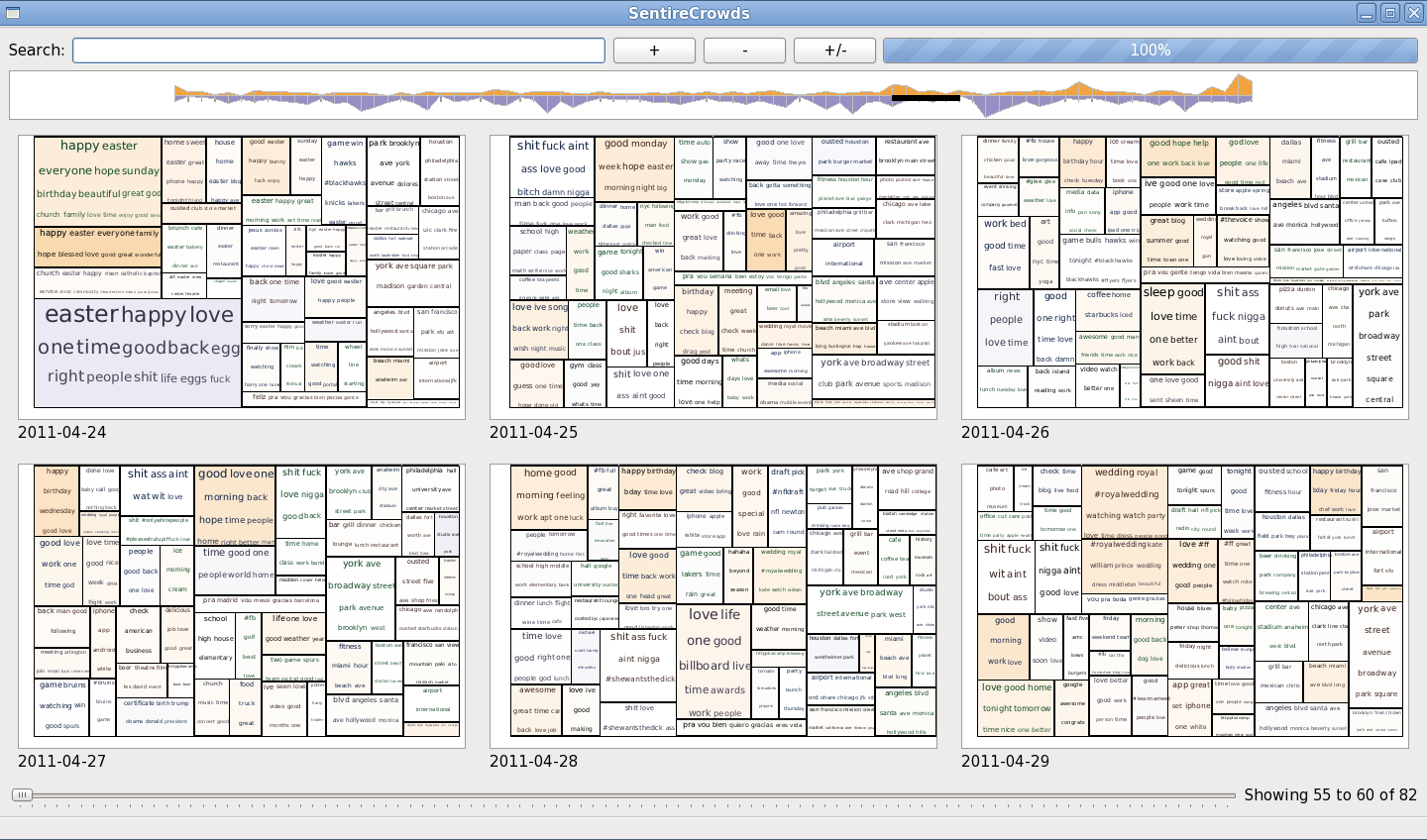}}
\caption {Case studies for {\tt US Cities} data set. {\bf (a)} Microblog activity around the day that Osama bin Laden was killed, demonstrating the evolution of discussions around the event.  ThemeCrowds depicts the evolution of conversations around this event.  {\bf (b)} Twitter activity around the Easter holiday and the marriage of Prince William and Catherine Middleton.  SentireCrowds is used to depict topics of strong sentiment around this time.}
\label {figUSCities}
\end {figure*}

\begin {figure*}
\centering
\subfigure [Democratic National Convention - \hashtag{dnc2012} (ThemeCrowds)] {\label {dncMulti}\includegraphics[width=\linewidth] {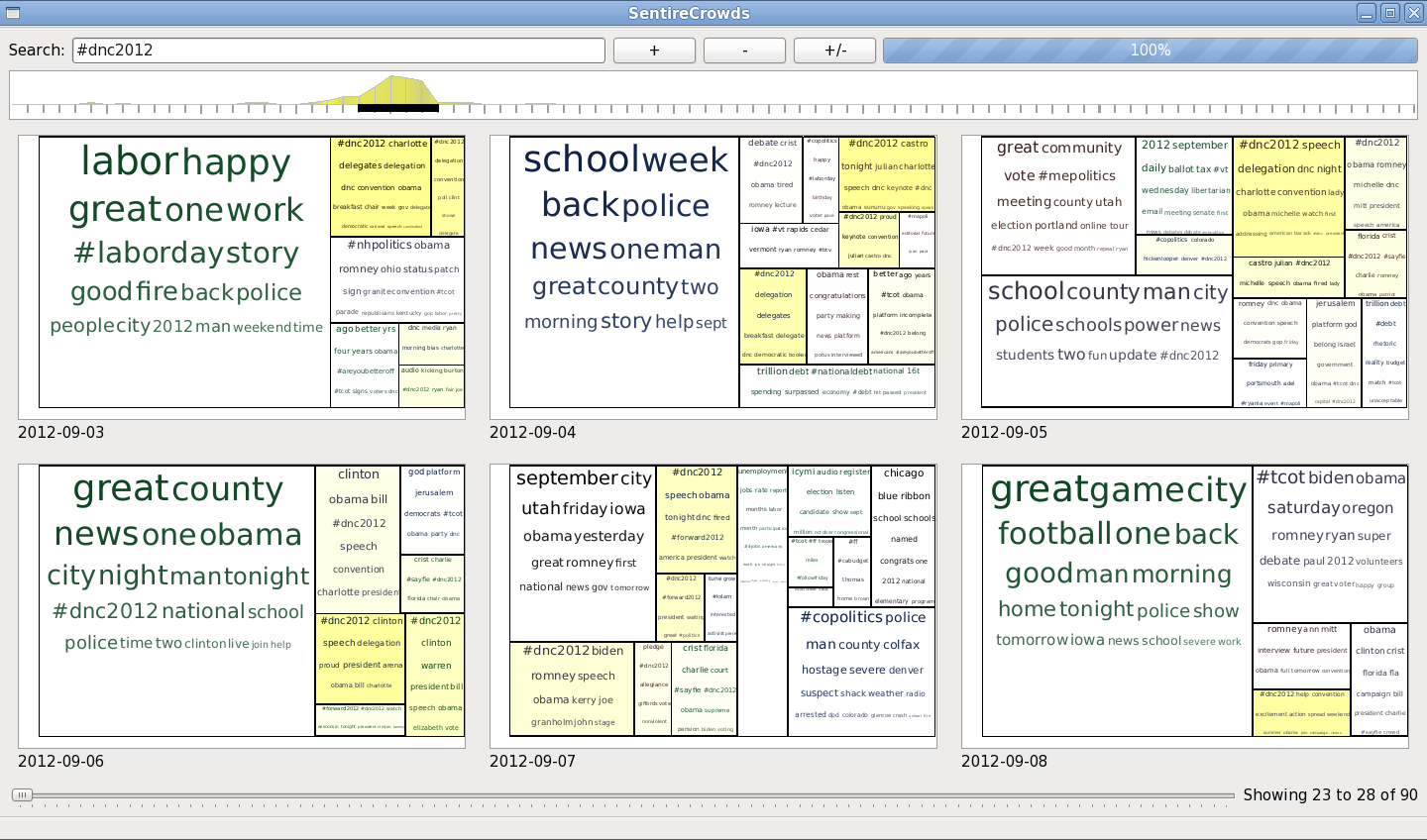}}
\vskip 1em
\subfigure [US Consulate Attack, Bomb Threat, Rosh Hashanah, and Constitution Anniversary (SentireCrowds)] {\label {consulateList}\includegraphics[width=\linewidth] {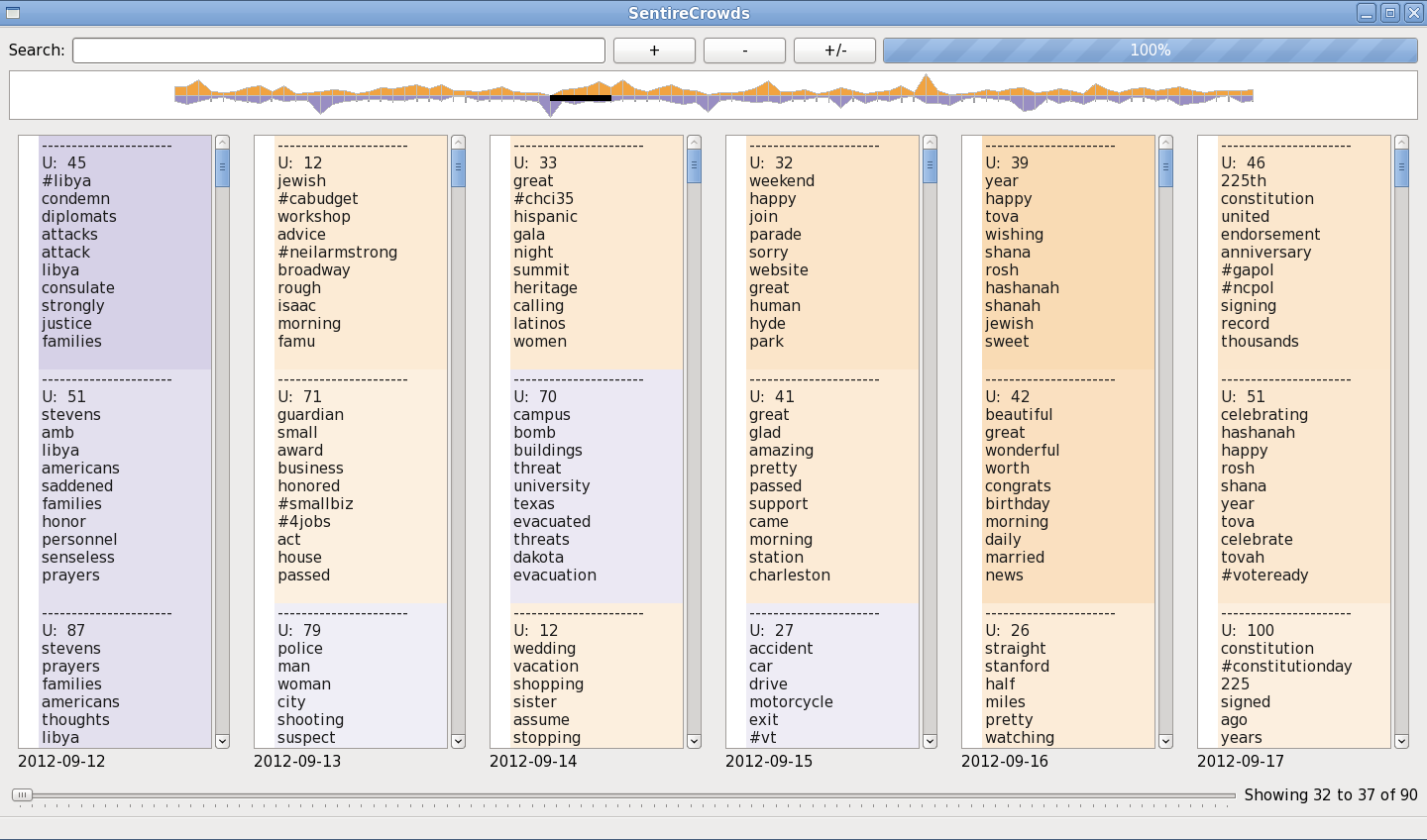}}
\caption {Case studies for {\tt Election~2012} data set. {\bf (a)} Progression of the Democratic National Convention (DNC 2012) by visualizing conversations enriched with \hashtag{dnc2012}.  ThemeCrowds depicts how the hash tag is used over time.  {\bf (b)} Sentiment around the US consulate attacks in September 2012.  SentireCrowds shows conversations of strong sentiment during this time.}
\label {figElection}
\end {figure*}

In this section, we describe some qualitative results when our visualization approaches were applied to the processed data sets described in the previous section.  Fig.~\ref {figUSCities} shows the results for {\tt US~Cities} while Fig.~\ref {figElection} shows the results for {\tt Election~2012}.  These figures provide a few examples of the many discussions visible using our technique around certain events occurring during these times.

In Fig.~\ref {osamaList}, we show the time period around the death of Osama bin Laden using the list version of the ThemeCrowds interface.  To generate this visualization, we searched for the key term {\it Osama}.  First reports of his death emerge on May 1st, 2011 locally in the US as he was killed in the early morning hours of May 2nd in Pakistan.  Three primary themes of discussion are visible on this day.  Firstly, a large number of users simply report the death and express some relief ({\it finally} is a high ranking term).  In the second discussion, the focus is more on the president accomplishing this feat as {\it Obama} and {\it mission} feature much more prominently.  Finally, the third discussion is the circulation of a joke on Twitter where users said that Donald Trump was requesting to see Osama bin Laden's death certificate.  The reports of bin Laden's death continue into May 2nd, but another important theme emerges as seen in the second discussion on this day.  Users begin to report that Osama bin Laden's body was buried at sea.  Finally, on May 4th, we see a new discussion topic around Osama bin Laden where users discuss some leaked photos.  This visualization system helps users explore what people are saying about a given topic, like Osama bin Laden, and how that discussion evolves over time.

\begin {figure}[!t]
\centering
\subfigure [Easter] {\label {figZoomEaster}\includegraphics[width=0.48\linewidth] {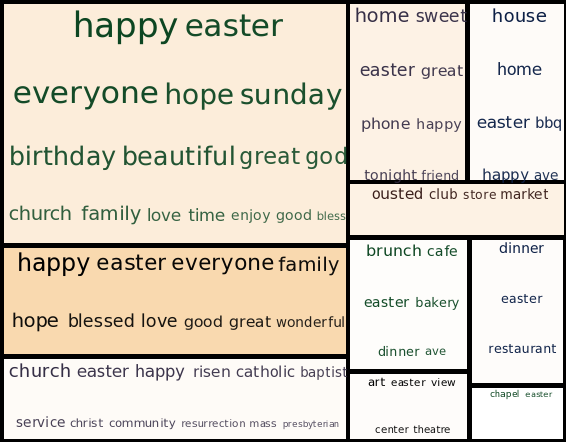}}
\hskip 4em
\subfigure [Wedding] {\label {figZoomRoyalWedding}\includegraphics[width=0.25\linewidth] {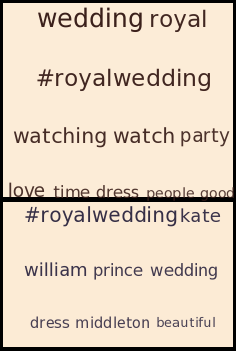}}
\caption {Close-up of {\tt US~Cities} discussions in SentireCrowds (Fig.~\ref {easterRoyalMulti}).  {\bf (a)} Close-up of discussions around Easter on April 24th, 2011.  {\bf (b)} Close-up of discussions around the British royal wedding on April 29th, 2011.}
\label {figZoomEasterRoyalWedding}
\end {figure}

Fig.~\ref {easterRoyalMulti} shows discussions of strong sentiment around the dates of April 24th - April 29th, 2011 using the multilevel tag cloud version of SentireCrowds.  During this week, both Easter and the British Royal Wedding were prominent positive discussions.  In the upper left hand corner on April 24th, we see a number of discussions about wishes and what people are doing for Easter (close-up Fig.~\ref {figZoomEaster}).  As the decomposition takes topic into account, notice that most of these discussions are situated in the upper left hand corner of the display as they share a least common ancestor that is not that far above them in the hierarchy.  The group of users in the far upper left seem to primarily be sharing Easter wishes.  Directly below this group two discussions are more focused on family and religion.  To the right of this group, a discussion focuses on being home for the holiday.  On April 29th, there are a few groups of users talking about the wedding of Prince William and Catherine Middleton (close-up Fig.~\ref {figZoomRoyalWedding}).  The first of these discussions mentions that the users are watching the event and parties around the event.  The second discussion, directly below the first, are messages primarily complementing Catherine Middleton's dress.

\begin {figure}[!t]
\centering
\includegraphics[width=0.55\linewidth] {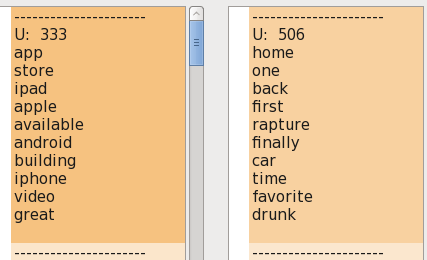}
\caption {Close-up of {\tt US~Cities} but using the list interface of SentireCrowds.  These groups of discussions are those of strongest sentiment on the dates of May 20th and 21st, 2011.  On the left, discussions are actually a spam campaign and on the right are satirical discussions around the prediction of the rapture.}
\label {figSpamRapture}
\end {figure}

In Fig.~\ref {easterRoyalMulti} notice the very strong positive spike in sentiment at the end of the time series (around May 21st, 2011).  Seemingly, there was no major event occurring in the media on this day.  A close-up of this anomaly is shown using the list interface of SentireCrowds in Fig.~\ref {figSpamRapture}.  It turns out that there were two anomalous events in our Twitter data set that occurred during this time.  The first event corresponded to a spam campaign, involving great deals on information technology devices with words such as {\it app store}, {\it ipad}, {\it android}, {\it iphone}, and {\it available} featuring prominently.  The second is discussion around the prediction of the end of the world by American Christian radio broadcaster Harold Camping\footnote {http://www.bbc.co.uk/news/world-us-canada-13489641}.  A substantial number of ironic or satirical comments surrounding the story (e.g. {\it ``Pre rapture party. Best idea ever''â}, {\it ``I can't think of a rapture joke, I'm not worrying, it's not the end of the world''}) seem to contribute to this spike in sentiment.

Fig.~\ref {dncMulti} shows how discussions around the Democratic National Convention progressed between September 3rd - 8th, 2012 using the multilevel tag cloud version of the ThemeCrowds interface.   To generate this visualization, we searched for the key term {\it \#dnc2012}.  The visualization demonstrates how ThemeCrowds can present the progression of an event over time.  The first visible day shows some general discussion about the convention and some remarks related to Clint Eastwood's ``empty chair'' speech at the Republican National Convention which took place earlier in the month\footnote{\url{http://www.bbc.co.uk/news/entertainment-arts-19434705}}.  Discussion then shifts to the keynote speech of {\it Julian Castro} which occurred on September 4th.  The day after, the reaction to the speech of {\it Michelle Obama} is quite prominent.  On September 6th, discussion about {\it Bill Clinton} and {\it Elizabeth Warren}'s speeches emerge.  Discussion about the president's acceptance of the nomination occurs on the 7th and general discussion persists on the 8th.  The example demonstrates the ability of this technique to visualize how the language around a given topic can evolve during a live event, such as DNC 2012. 

Fig.~\ref {consulateList} shows sentiment around discussions about events that occurred during the week of September 12th - 17th, 2012 using the list version of SentireCrowds.  During this time period, a number of events occurred to which users of Twitter reacted strongly.  On September 12th, users discuss in a very negative light the attacks on the US consulate.  The discussion of strongest sentiment condemns the attacks and contains some calls for justice.  The second and third discussions primarily express thoughts and prayers for all those affected by the attacks.  On September 14th, the University of Texas at Austen and North Dakota State University received bomb threats and were required to evacuate their campuses\footnote {\url{http://www.bbc.co.uk/news/world-us-canada-19602986}}.  Reactions of negative sentiment to these events are seen in the second discussion on this day.  Later in the week, the discussions of strong sentiment are positive.  On the 16th, discussions about Rosh Hashanah, the Jewish new year becomes a strong positive discussion.  The following day, there is positive discussion about Constitution Day in the United States.  By visualizing groups of users speak about similar topics and selecting antichains in the data based on sentiment, the technique is able to summarize how the most positive and negative discussions in the data evolve over time.

%!TEX root = twittercrowds-tr.tex
\section{User Study}
\label{sec:eval2}

To test the effectiveness of multilevel tag clouds and lists on typical social media visualization tasks involving topics and sentiment, we performed a within-participant experiment.  We employed a 2 interface (multilevel tag clouds vs lists) $\times$ 2 data set ({\tt US Cities} and {\tt Election 2012} as in \refsec{sec:data}) $\times$ 4 question $\times$ 2 repetition design.  In this section, we present the design and results of this experiment.

\subsection {Tasks}

The tasks selected for this experiment were derived from visualization tasks related to topic and sentiment  visualization in social media.  Two of the four questions are topic-based and the remaining two are sentiment-based questions.  For questions based on topic, the decomposition of the time series was automatically computed by the test setup and the yellow/white colouring scheme used throughout the paper indicated clusters relevant to the question.  For questions based on sentiment, the decomposition by sentiment was automatically computed by the test setup using the tan/purple colouring scheme used throughout the paper.  Dates of the daily time steps were obfuscated ({\it $t1,t2,t3,\dots$} instead of {\it March 16th, March 17th, March 18th,\dots}) in order to prevent prior knowledge of the events from impacting answers.  For all tasks, the interface was also automatically scrolled to the time window relevant to the question.  We chose automatic scrolling to eliminate the confound of using the scented widget through the time series and exclusively test the representation of the microblogging data.  The discussion clusters on both interfaces were limited to ten words so that both interfaces contained the same information content.  In all cases, answers were multiple choice and entered using radio buttons.  A button at the bottom of the screen was used in order to confirm answers.  

The first task asks participants to find the first instance when a given topic is discussed in a particular light.  The form of these questions was {\it Given a topic X, when is the first instance of discussion Y}.  This type of question tests the ability of participants to understand what people are saying about a given topic in a way that has a unique answer.  For example:

\begin {enumerate}
\setcounter {enumi}{0}
\item On {\bf which day} is red cross donations for japan first discussed?
\end {enumerate}

Any of the six visible time steps was offered as a possible answer, and the correct answer was the first day in the data where the above statement was true.

The second task examines how participants are able to use the interfaces to follow the progression of a discussion around a given topic.  The form of these questions was {\it On what days do the following discussions take place about X:  event A, event B, and event C}.  This type of question tests the participant's ability to follow the change in discussion about a topic over time in a way that has a unique answer.  For example:

\begin {enumerate}
\setcounter {enumi}{1}
\item On what days do the following {\bf discussions} take place about hurricane Sandy:  discussions about preparations begin, evacuations are discussed, a ship is declared missing off the east coast.
\end {enumerate}

Participants were presented with six possible sequences of three (not necessarily consecutive) days as answers (for example {\it $t2,t4,t6$}).  To disambiguate the case where discussions occur over multiple days, participants were asked to pick the earliest day.

The third task tested the ability of the interface to communicate areas of maximum positive or negative sentiment.  In this question, participants were asked one of the following two forms of the same question:

\begin {enumerate}
\setcounter {enumi}{2}
\item What is the {\bf most positive}/{\bf most negative} discussion displayed in this time window?
\end {enumerate}

Possible answers consisted of six keywords, one of which described the most positive or most negative discussion present in the time window.  The most appropriate keyword was the correct answer.

The final question asks participants to find the first mention of a topic in a positive or negative light.  This question is similar to the first question but deals with sentiment rather than topic.  The form of these questions was {\it When do the microbloggers first begin to talk positively/negatively about X?}  For example:

\begin {enumerate}
\setcounter {enumi}{3}
\item When do the microbloggers {\bf first} begin to talk {\bf positively} about \hashtag{earthday}?
\end {enumerate}

Any of the six visible time steps was a possible answer, and the correct answer was the first of the six time steps visible in the time window that was negative/positive and about the indicated topic.

\subsection {Experimental Design}
As multilevel tag clouds and lists are sufficiently different interfaces, the experiment was divided into two counterbalanced blocks of 16 questions each.  During each block, the four questions on each data set were asked exactly twice ($4 \times 2 \time 2 = 16$).  To overcome the learning effect, these 16 tasks were randomized individually per participant and were prefixed with four practice tasks.  The practice tasks presented all questions exactly once and both data sets twice.  These tasks were discarded from the experimental results, and participants were not made aware that they did not form part of the experiment.  Therefore, for each interface, participants performed a total of 20 tasks.  These tasks were divided into two blocks of 10 questions, between which participants could take a short break.

We counterbalanced between participants by presenting multilevel tag clouds first to even-numbered participants and lists first to odd-numbered participants.  The participant could take a short break between experimental conditions.  The experiment required participants to answer all tasks under one interface first, followed by all questions on the second interface.  Therefore, any cognitive shift required to move between interfaces occurred once.  Before the start of each interface block, participants had a demonstration session which introduced them to the experimental interfaces.  During this session, the participants could ask questions, find out about the experimental tasks and learn how to find the answers to the questions.

Overall, there were 22 participants used in the final results.  Participants were drawn from members of the Complex and Adaptive Systems Laboratory at University College Dublin.

\subsection {Results}

We present the results for our experiment comparing Multilevel Tag Clouds (Mult.) to Lists (List) in terms of response times and error rates.  In all of our statistics, a Shapiro-Wilk test, with a significance level of $\alpha = 0.05$, was used to determine whether or not the data was normally distributed.  We found that, in all cases, at least one distribution was not normal for both response time and error rate. As a consequence, we used an exact Wilcoxon signed rank at a significance level of $\alpha = 0.05$.  When we divided the data by target level, we applied a Bonferroni correction, thus reducing the significance level to $\alpha = 0.0125$.  In all bar charts, black lines connect pairs of bars with significant differences.  Below each bar, mean and median are indicated and separated by a hyphen.  The standard error is indicated on the top of each bar.

\subsubsection {Overall}
\begin {figure}[!t]
\centering
\subfigure [Response Time] {\label {allTime}\includegraphics[width=0.38\linewidth] {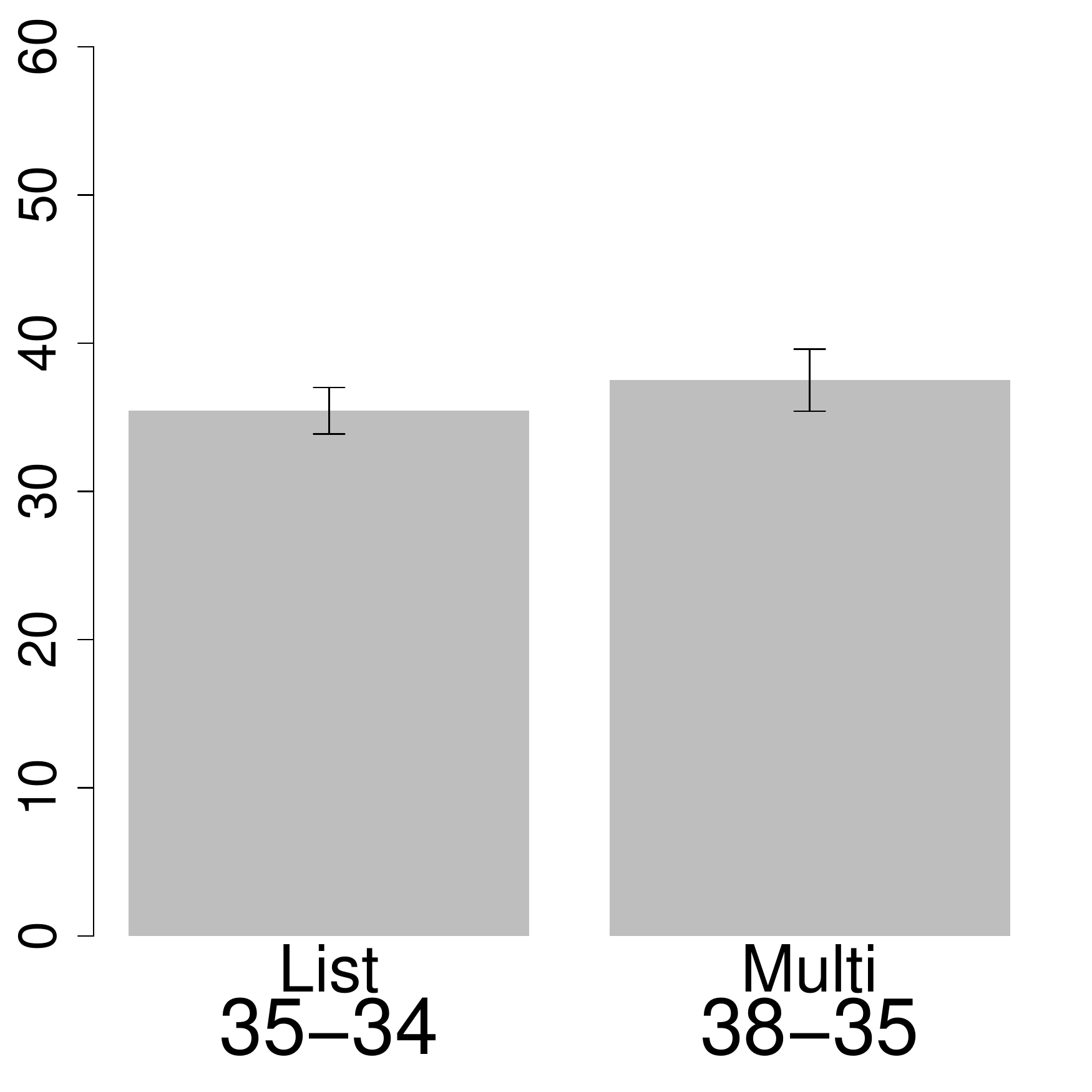}}
\hskip 3em
\subfigure [Error Rate] {\label {allError}\includegraphics[width=0.38\linewidth] {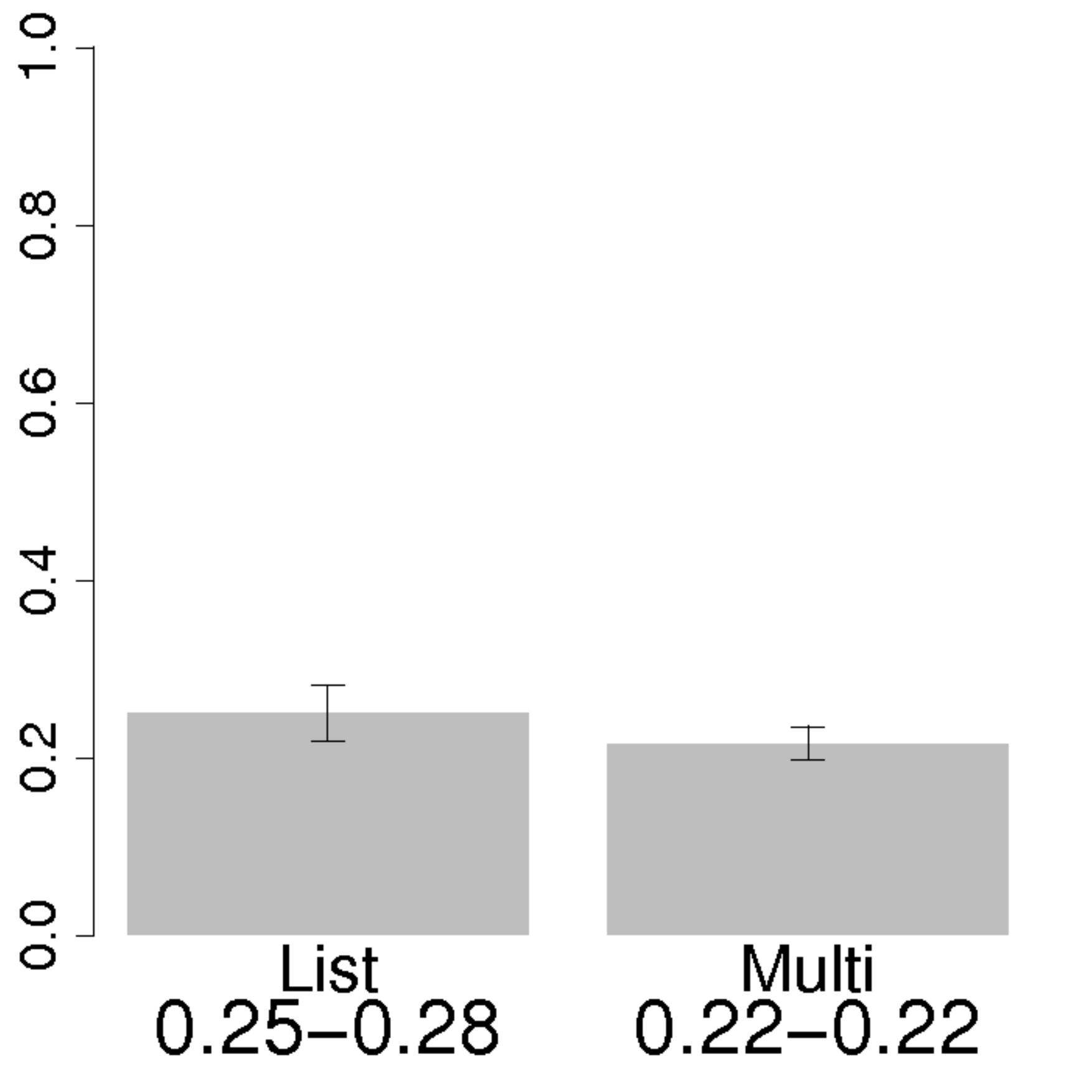}}
\caption {Response time in seconds and error rate for all questions, comparing Multilevel Tag Clouds ({\it Multi}) and Lists ({\it List}).}
\label {resultOverall}
\end {figure}
Fig.~\ref {resultOverall} shows the mean error and response time results when comparing multilevel tag clouds to lists.  No significant difference was found either in terms of error rate or response time.  This result is expected as our tasks are varied.

\subsubsection {Divided by Question}
\begin {figure*}[!t]
\centering
\begin {tabular} {|c||c||c||c|}
\hline
{\bf Question 1} & {\bf Question 2} & {\bf Question 3} & {\bf Question 4}\\
\hline\hline
\includegraphics[width=0.22\linewidth] {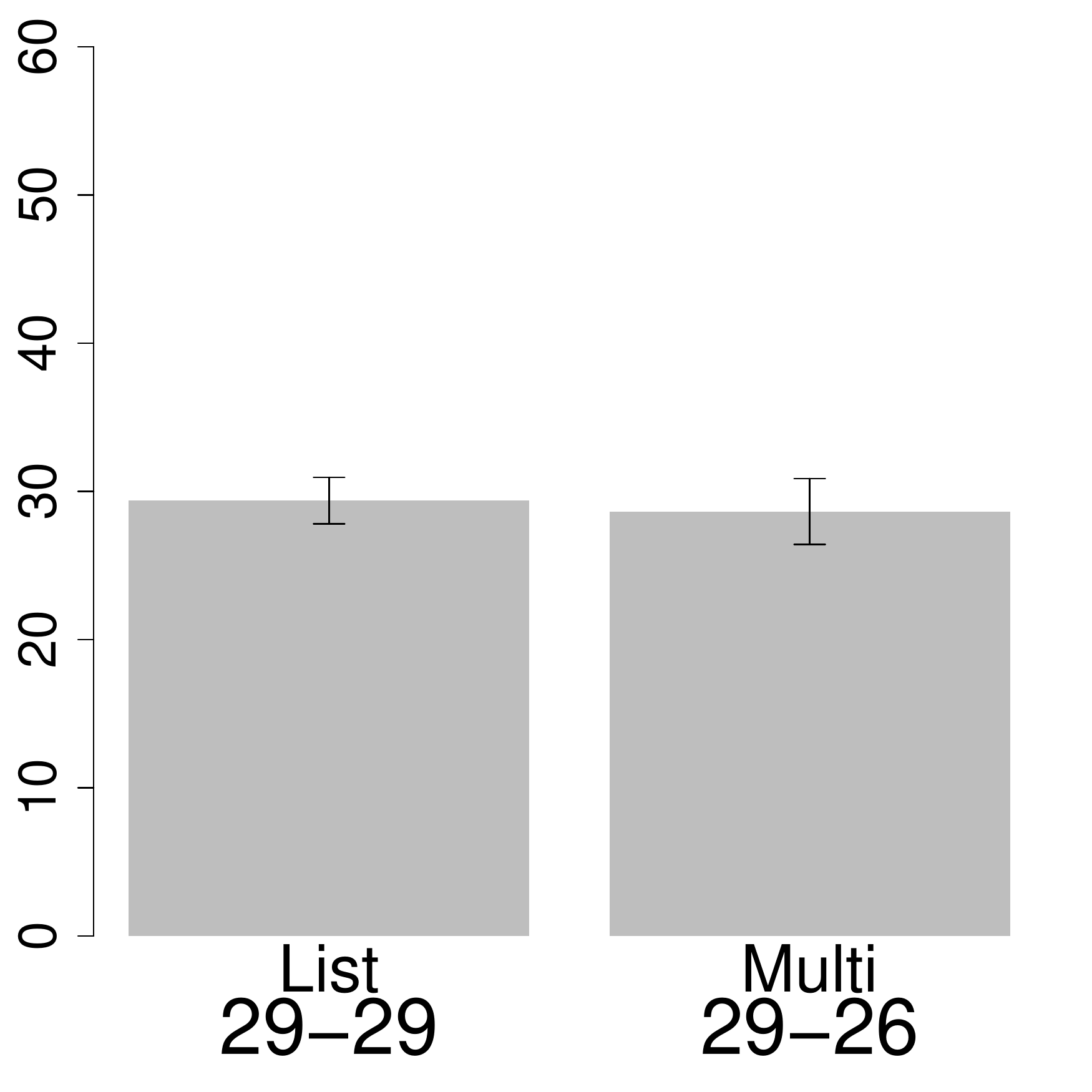} &
\includegraphics[width=0.22\linewidth] {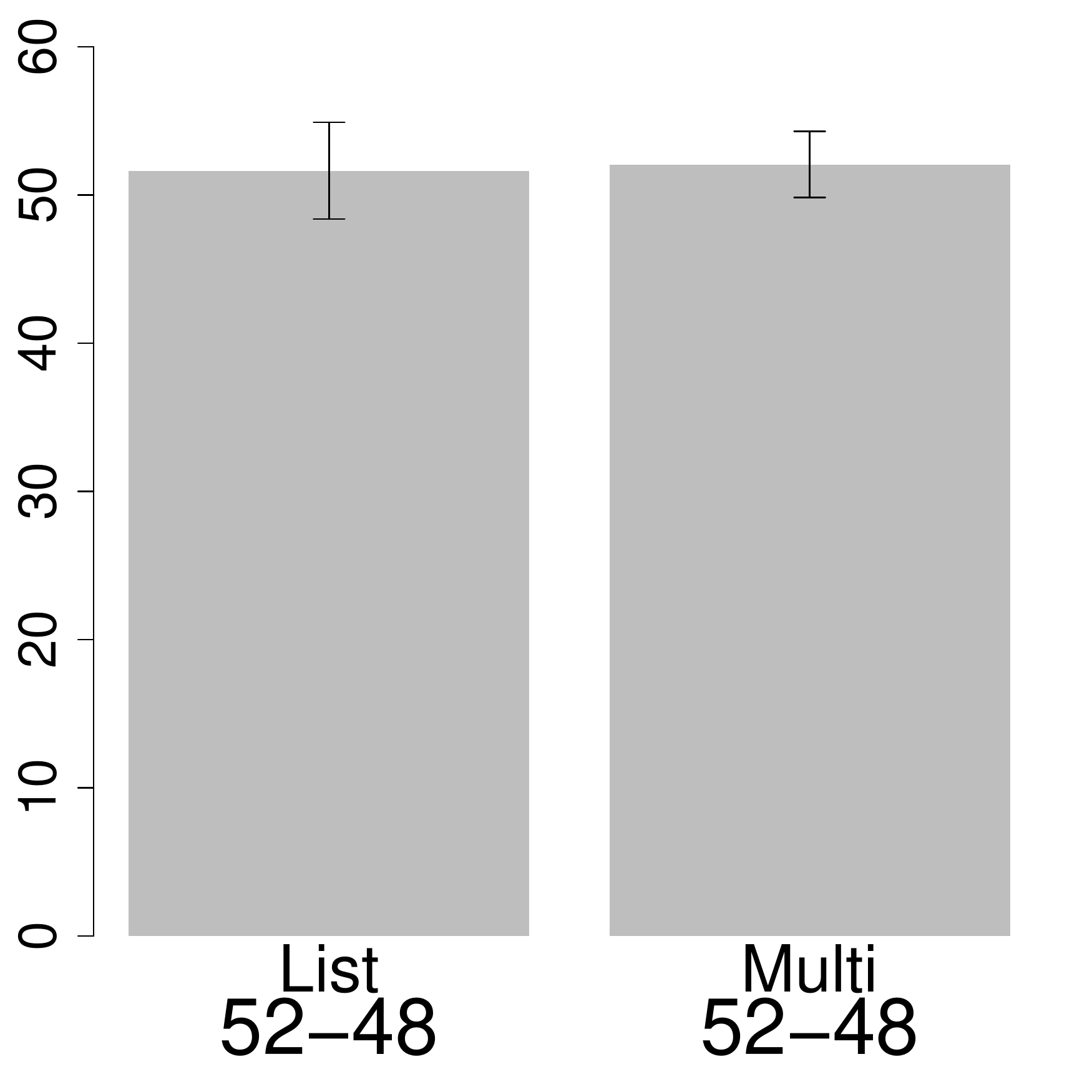} &
\includegraphics[width=0.22\linewidth] {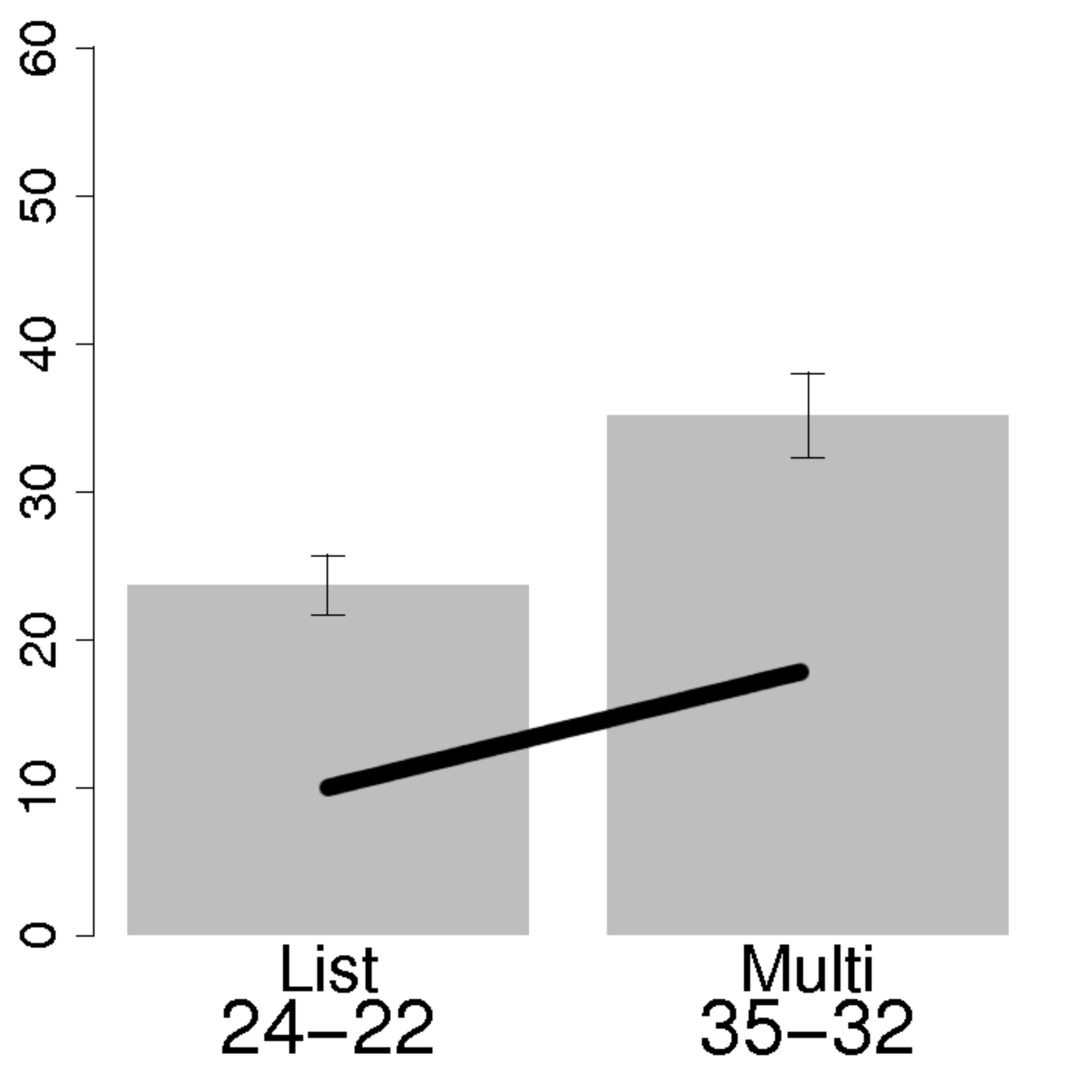} &
\includegraphics[width=0.22\linewidth] {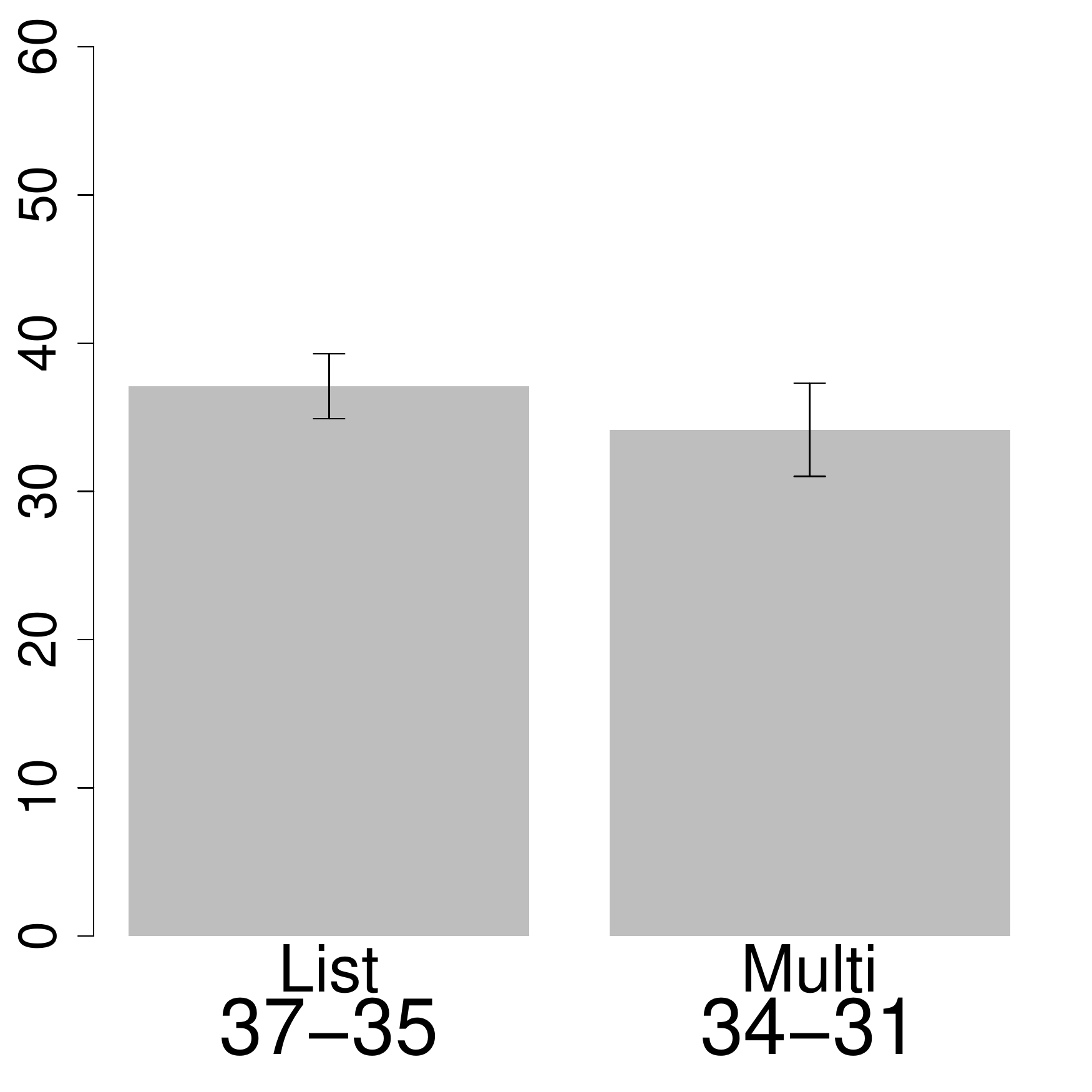} \\ 
\hline
\multicolumn{4}{|c|} {{\bf Response Time (Seconds)}}\\
\hline\hline
\includegraphics[width=0.22\linewidth] {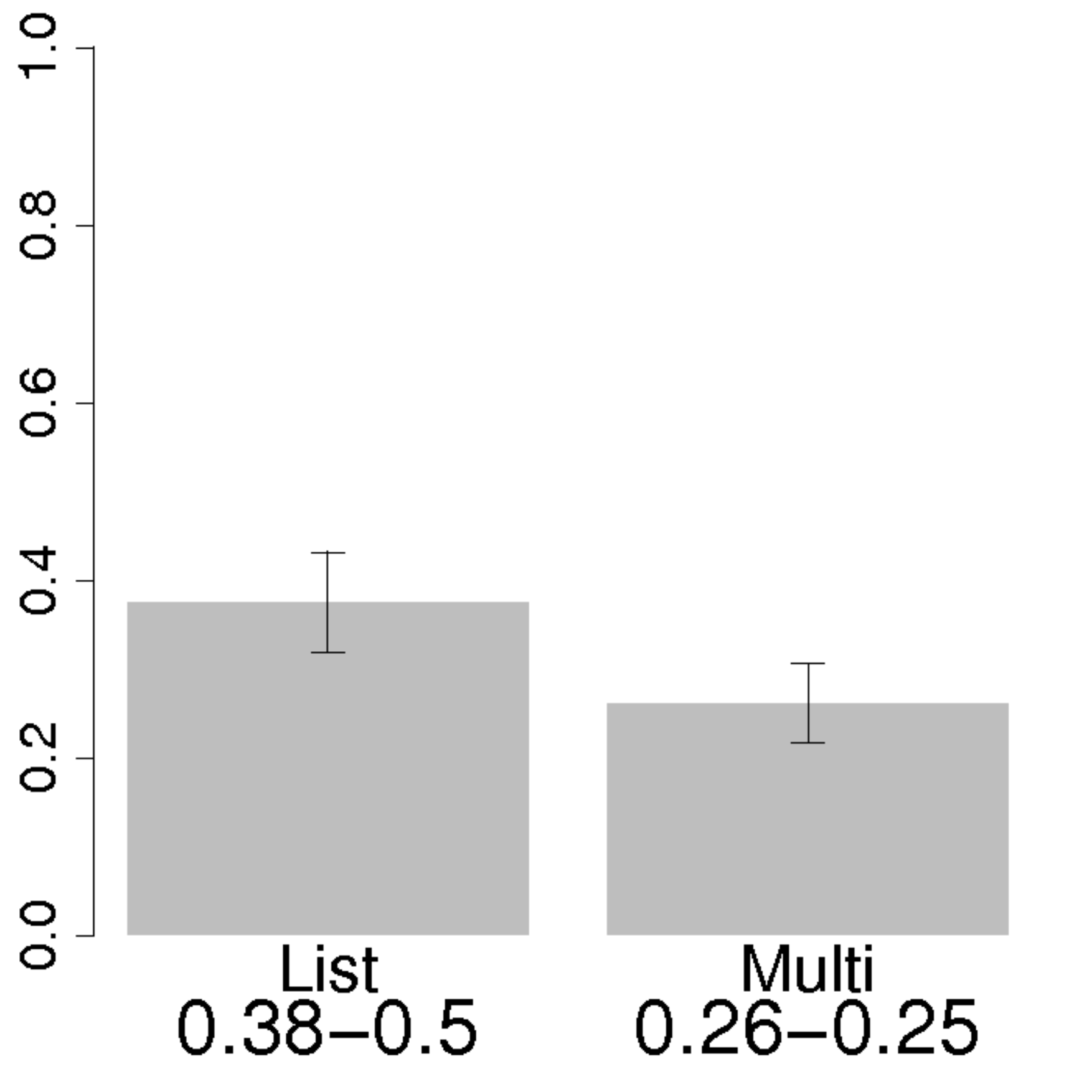} &
\includegraphics[width=0.22\linewidth] {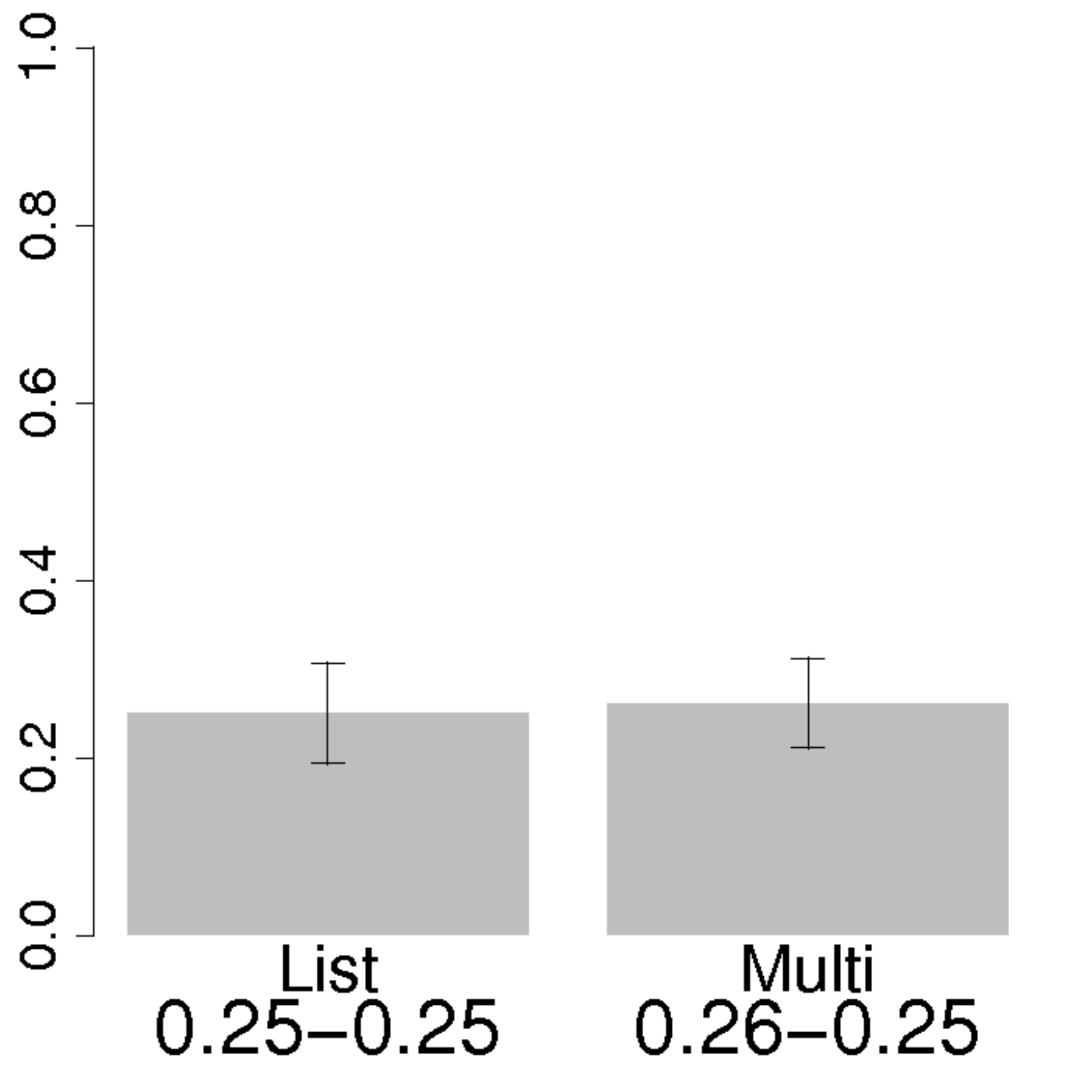} &
\includegraphics[width=0.22\linewidth] {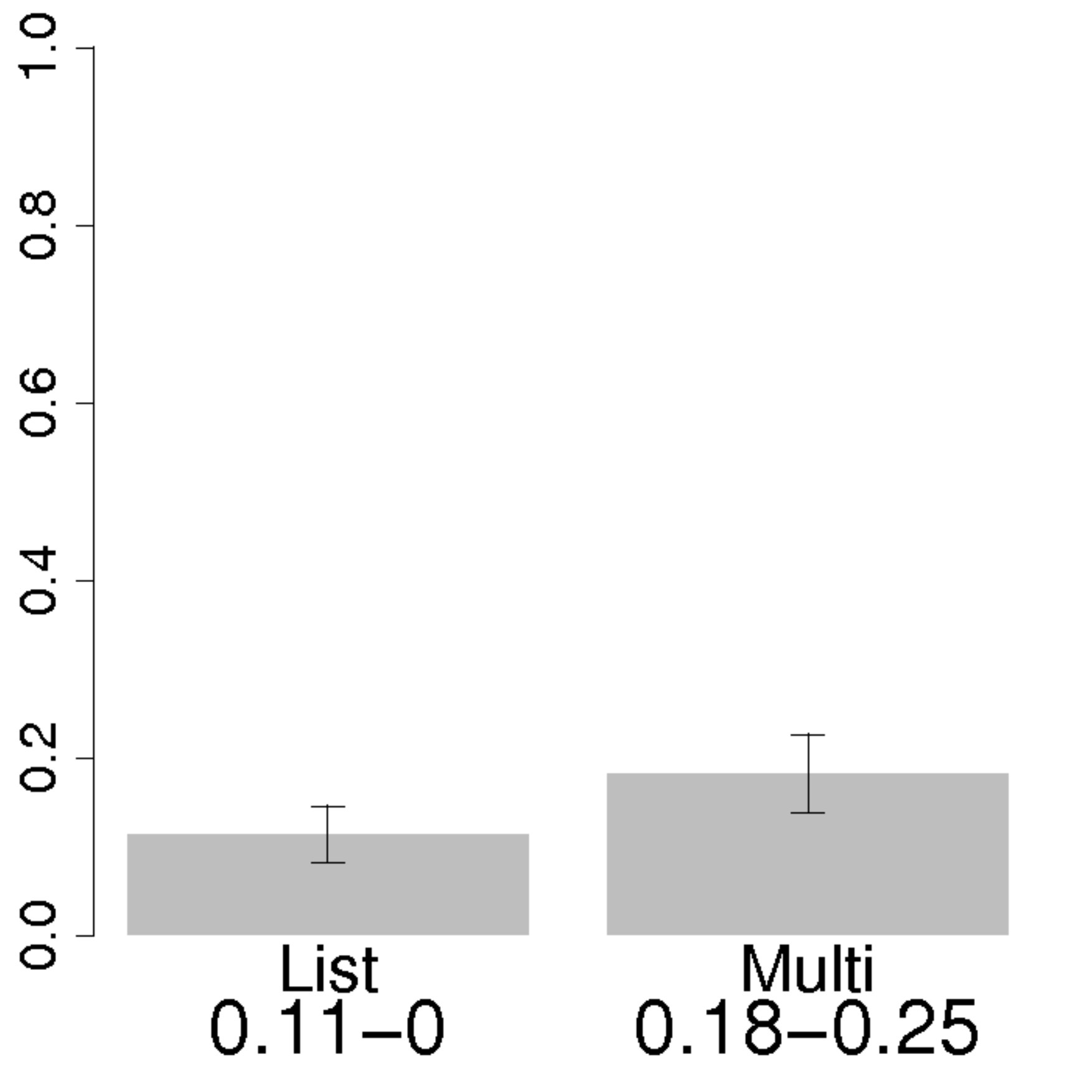} &
\includegraphics[width=0.22\linewidth] {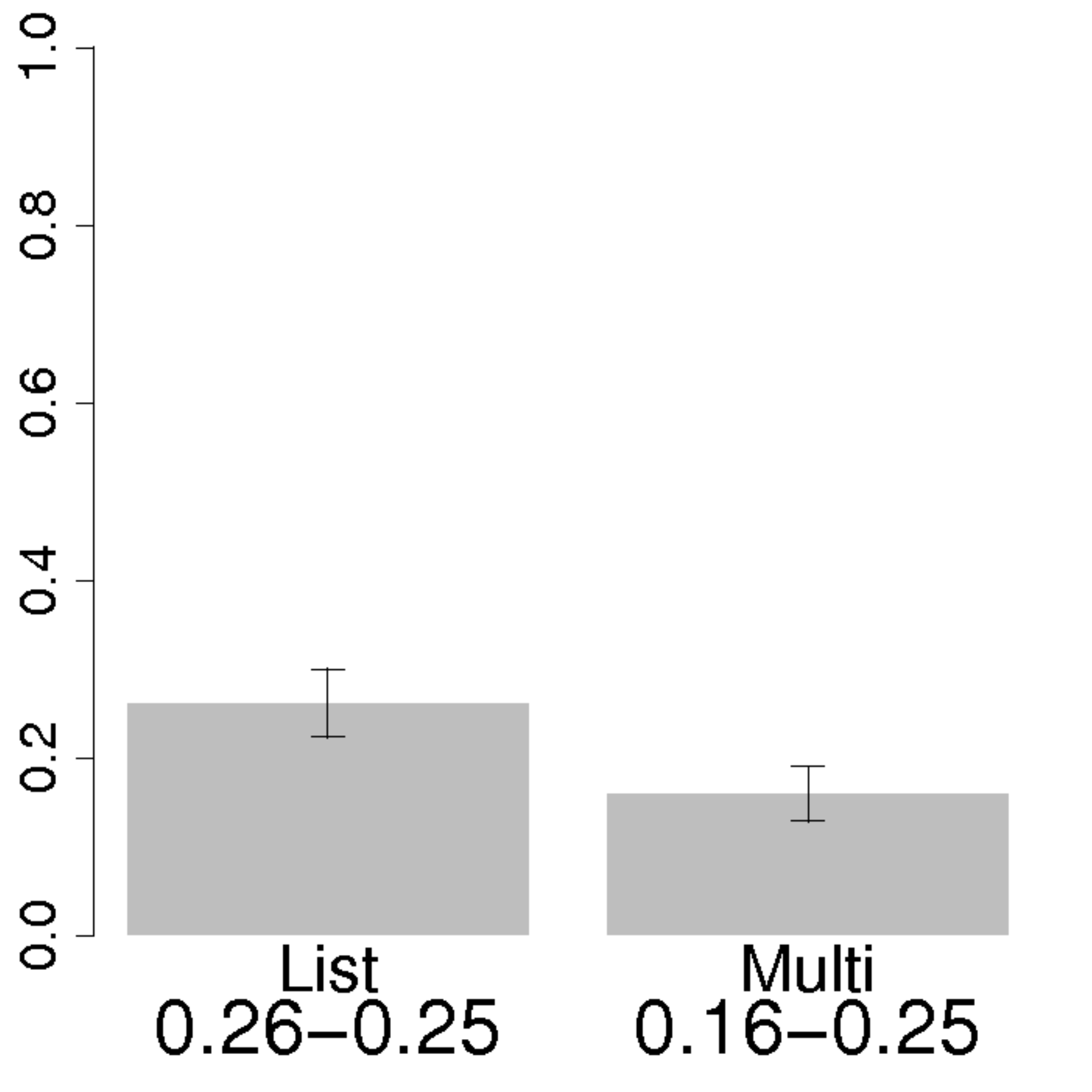} \\
\hline
\multicolumn{4}{|c|} {{\bf Error Rate}}\\
\hline
\end {tabular}
\caption {Response time and error rate when divided by question for Multilevel Tag Clouds ({\it Multi}) and Lists ({\it List}).}
\label {divQuestion}
\end {figure*}

We subsequently divided the data by question.  Fig.~\ref {divQuestion} shows the mean error and response time results when comparing multilevel tag clouds to lists after this division.  When we divided by question, we found the following results:

\begin {itemize}
\item {\bf Minimum/Maximum Sentiment}:
\begin {itemize}
\item Lists was significantly faster than multilevel tag clouds (Mult.: 35.1s; Lists: 23.3s; $p=0.00043$).
\item No significant difference in terms of error rate.
\end {itemize}
\item {\bf All Other Questions}:
\begin {itemize}  
\item No significant difference in terms of response time or error rate.
\end {itemize}
\end {itemize}

\subsection {Preference Data}

In this section, we present the preference data for the experiment.  Table~\ref{prefDataTable} indicates, per overall and per question, how many participants preferred each interface.  For most of the questions and overall, multilevel tag clouds were preferred to lists.  In the case of Q2, however, the result was much closer but still in favour of multilevel tag clouds.   

\begin {table}[!h]
\centering
\caption {Preference data overall and per question.  Number of participants that preferred the indicated interface present in the table cell.}
\begin {tabular} {|l||r|r|}
\hline
& {\bf Multilevel} & {\bf Lists}\\
\hline\hline
{\bf Overall} & 13 & 9\\
\hline
{\bf Q1} & 15 & 7\\
{\bf Q2} & 12 & 10\\
{\bf Q3} & 15 & 7\\
{\bf Q4} & 14 & 8\\
\hline
\end {tabular}
\label {prefDataTable}
\end {table}

\subsection {Discussion}
\label {secDiscussion}
Lists were significantly faster than multilevel tag clouds for tasks involving finding areas of maximum sentiment.  For all the questions involving sentiment in our experiment, the lists were sorted by sentiment strength and placed strong sentiment always near the top of the list.  As the main advantage of lists stems directly from their ordering, visual search time was reduced, leading to a significant improvement in response time.  Thus, our findings provide evidence that lists are more effective for this analysis task.

If such an ordering is unknown, it seems that participants needed to resort to visual search and scrolling, causing performance to be comparable.  An optimal ordering is not entirely obvious for the remaining questions.  All questions involved the type of language used around a given topic, and the user is not aware of what is being said about the topic in advance.  Therefore, keywords cannot be entered beforehand, making sorting difficult.  Thus, the two representations have similar performance if an ordering is not known beforehand, but lists perform best if an appropriate ordering for the task is available.  This result is consistent with previous experiments comparing lists to tag clouds, where alphabetical order~\cite {07Halvey} and word frequency~\cite {07Rivadeneira} seemed to be clear orderings for the tasks tested in these experiments.  

Multilevel tag clouds were preferred overall and in all questions -- in many cases overwhelmingly so.  This preference for multilevel tag clouds was due to the fact that the interface was more engaging, colourful, and fun to use.  One participant explicitly commented on this fact, saying that he enjoyed the multilevel tag clouds more but was doubtful that he performed better with them.  This argument is reminiscent of arguments for animation as a visualization and learning tool as animations are more attractive, motivating, and fun.  However, they can frequently take more time, so they have a cost~\cite {02Tversky}.  The ordering result and survey data explicitly support a very similar argument to that of Tversky {\it et al.}~\cite {02Tversky}, when comparing tag clouds and list interfaces for visualizing microblogging data.  In the cases when an ordering is not so clear and performance between the two interfaces is similar, user preference potentially plays a more important role.

%!TEX root = tvcg2013.tex

\section{Conclusion}
\label{sec:conc}

In this paper, we present extended descriptions of the ThemeCrowds~\cite {11ThemeCrowds} and SentireCrowds~\cite {Brew2011} visualization interfaces and introduce list equivalents of both.  We demonstrate the capabilities of these techniques on microblogging data of realistic size, consisting of tens of millions of tweets.  In order to evaluate the effectiveness of these approaches, we performed a user study involving tasks these interfaces were designed to answer.  In our user study, we found that the list interface was faster for questions which involved determining areas of maximum positive or negative sentiment.  In terms of user preference, multilevel tag clouds were found to be more enjoyable to use.  Our experiment gives support to lists as effective presentation method for data structured in this way, if a suitable ordering of the list is available for the task.  This result is consistent with other experiments in the literature on tag clouds~\cite {07Rivadeneira,07Halvey}.  However, both interfaces were usable and generally had low error rates for the data sets and tasks tested.  

The visualization techniques presented in this paper allows for scalable visualization of sentiment and topics in tens of millions of tweets.  These interfaces scale by leveraging tools in unsupervised learning, where data of this size is common, and adapting information visualization methods to this problem.  This work provides a first step in supporting the tasks that members of industry and researchers in the social sciences would like to investigate in microblogging data.  However, the user community, especially those in industry, would like results to be more real time for decision making.  In future work, we hope to adapt some of these techniques in order to reduce the time required from data collection to visualization.  Possible methods include approximate clustering or adapting streaming methods for data processing to our analysis and visualization pipeline.

%-----------------------------------------------------

\section*{Acknowledgements}
\noindent 
This research was supported by Science Foundation Ireland Grant 08/SRC/I1407 (Clique: Graph and Network Analysis Cluster).

%%%%%%%%%%%%%%%%%%%%%%%%%%%%%%%%%%%%%%%%%%%%%%%%%%%%%%%%%%%%%%%%%%%%%%
\small
\bibliographystyle{abbrv}
\bibliography{twittercrowds-tr}

\end{document}